\documentclass[structabstract]{aa}  
\usepackage{graphicx}
\usepackage{txfonts}
\begin{document}
 \title{The modulation of SiO maser polarization by Jovian planets}

   \author{H.W. Wiesemeyer\thanks{On leave to Instituto de Radioastronom\'ia
           Milim\'etrica, Granada, Spain}}

   \institute{Institut de Radioastronomie Millim\'etrique,
              300 rue de la Piscine, Domaine Universitaire,
	      F-38406 Saint Martin d'H\`eres \\
              \email{wiesemey@iram.fr}}

   \date{Received ; accepted }

 
  \abstract
   {}
   {Searching for planets in the atmosphere of AGB stars is difficult,
    due to confusion with the stellar wind and pulsations. The aim here
    is to provide a complementary strategy for planet searches in such a
    dense environment.}
   {The polarization properties of SiO masers, especially their circular
    polarization, are, under certain conditions, good tracers of rapid
    magnetospheric events. A Jovian planet with a magnetosphere whose dipole
    axis is misaligned with its rotation axis naturally provides such
    conditions. Here I present several models showing that the polarization will
    be periodically modulated.}
   {The linear and circular polarization of an SiO maser in a planetary magnetosphere
    is modulated by the precessing dipole component of the latter. The effect is
    measurable in saturated masers, while unsaturated masers only exhibit weak changes,
    because of dilution effects, and because the circular polarization there stems from the
    Zeeman effect making it as weak as for thermal radiation. The situation would
    change if anisotropic pump- and loss-rates were included, which would increase the
    fractional linear and, via magnetorotation, the circular polarization of the modulation.}
   {Single-dish monitoring with a dense enough time sampling and a carefully
    calibrated polarimeter, in combination with VLBI observations, are suited to detecting
    and locating a periodic modulation of the circular maser polarization due to a
    precessing Jovian magnetosphere. The phenomenon will be rare, because a favorable
    arrangement of maser and magnetosphere is needed. Otherwise the polarization may be
    below the detection threshold, especially if the maser is unsaturated. Though exhibiting
    a qualitatively similar modulation, linear polarization is likely to suffer more from
    confusion due to dilution of the magnetosphere within the maser cross section, even
    in VLBI observations.}

   \titlerunning{}
   \maketitle
%

\section{Introduction}

    Most SiO masers are hosted by the extended atmosphere of evolved stars of
    low- to intermediate mass in the upper part of the asymptotic giant branch,
    before they evolve towards central stars of planetary nebulae (for a review see
    e.g., Herwig 2005). Under certain conditions, these masers trace the magnetic
    field by virtue of their polarization. After the first theoretical consideration
    of a Zeeman laser (Sargent et al. 1967), astronomical maser polarization has been
    extensively discussed, e.g., by Goldreich et al. (1977),
    Western \& Watson (1984, for linear polarization), Deguchi \& Watson (1986, for
    circular polarization), and Elitzur (1991). The models are compared by Gray (2003).
    The environment of SiO masers in AGB atmospheres is characterized by a dense wind
    driven by the stellar pulsations. Its interaction with planets engulfed in the stellar
    atmosphere has already been addressed by Struck-Marcell (1988) who investigated the
    hypothesis that SiO masers may form in the magnetosphere of Jovian planets. Struck
    et al. (2002, 2004) propose scenarios along the same lines, and model the dynamic
    effects of episodic accretion onto the planets. In a different context, the
    consequences of planets in the stellar atmosphere have been investigated by Soker
    (2001), who shows that Jovian and even Earth-like planets can efficiently spin-up
    the star after migration into the stellar atmosphere, such that it later may form an
    elliptical rather than a spherical planetary nebula.

    Here I propose to resume the idea of Struck-Marcell (1988) that SiO masers
    (though not all) may originate in Jovian magnetospheres, leading to
    circular maser polarization (Barvainis et al. 1987; Herpin et al. 2006),
    which may be as strong as $\sim 10\%$ of the Stokes I flux. I show that the densely
    sampled time series of all Stokes parameters yield a rare, but observable
    polarization signature characteristic of a precessing magnetosphere. Since 
    AGB stars are slow rotators, a precession period as fast as $\sim 10$\,h hints at a
    planetary magnetosphere. SiO masers are an ideal tool for sounding Jovian planets,
    since they arise in a zone extending from 1.5 to 7\,AU distance from the
    star where they are radiatively and collisionally pumped (Gray et al. 2009), before
    the SiO molecule becomes underabundant for maser action thanks to condensation in the dust envelope.
    The SiO maser zone thus corresponds to a distance from the star where the solar system
    harbors its giant gas planets (Jupiter is at about 5\,AU from the sun, and Saturn
    at about 10\,AU). A direct comparison between the solar sytem now and in the AGB
    phase is not valid, though, because planetary orbits may be dragged inwards by tidal
    friction, or expelled by the kinematical effect of stellar mass loss (Villaver \&
    Livio 2006, further references therein). Arguing with Soker (1996) that elliptical
    planetary nebulae formed from stars spun up by inward migration of a planet, it is 
    assumed here that the zone in question harbors Jovian planets, since 50\,\% of all
    planetary nebulae are not elliptical (Soker 2001).

    The first observational evidence of extrasolar planets around evolved stars was
    given by Silvotti et al. (2007) who indirectly detected a planet around a post-red
    giant star, measuring the pulse modulations of the variability of the latter. As for
    the planets detected around pulsars (with a similar method, measuring the period
    variations of the latter) by Wolszczan \& Frail (1992), it is not yet certain whether
    these planet-sized bodies survive the supernova explosion, or whether
    they formed in the debris disk forming in reponse to a ``fallback'' (Wang et al.
    2006) of ejected matter. Most extrasolar planets have so far been detected around
    main-sequence stars by radial velocity measurements (for a review see Santos 2008).
    As for the earliest phases of stellar evolution, a "hot Jupiter" has recently been
    discovered around the T Tauri star TW Hya (Setiawan et al. 2008). Around evolved
    stars, ten companions of substellar mass have been discovered to this day, the last one
    being the K0 giant HD17092 (Niedzielski et al. 2007, further references therein).
    There is no evidence yet from AGB stars, mainly because the strong, dense wind
    makes radial velocity measurements of the star difficult. The following models of
    maser polarization modulation due to precessing planetary magnetospheres are intended
    as a complement to exoplanet searches around AGB stars with different methods.


\section{Description of the model}

\subsection{Physical characteristics}

   SiO is non-paramagnetic, with a Zeeman splitting of
   $g\Omega [{\rm s^{-1}}] \sim 10^3 B [{\rm G}]$ (cf. Watson \& Wyld 2001).
   For a 1000~K gas and the magnetic fields considered here (of up to 10~G),
   the spectral linewidth is greater than the Zeeman splitting by a factor of
   $\sim 100$. This weak Zeeman splitting still produces a fractional circular
   polarization (hereafter $p_{\rm C}$) of $\sim 1\,\%$, which is observable for
   a carefully designed polarimeter (e.g., Thum et al. 2008). While linear
   polarization may not only be caused by magnetic fields, but also by anisotropic
   pumping in conjunction with the Hanle effect (Bujarrabal \& Rieu 1981;
   Western \& Watson 1984; Asensio Ramos et al. 2005), circular polarization is
   a tracer of magnetic flux density as long as the Zeeman splitting $g\Omega$ is much
   higher than the stimulated emission rate $R$ and the loss rate $\Gamma$. As soon as
   $R$ is close to $g\Omega$, the circular polarization will become intensity-dependent
   when the maser radiation saturates (Nedoluha \& Watson 1994), thus ceasing to be
   a reliable tracer of magnetic fields.

   To keep the model simple and to separate the predicted effects from
   those issued by models involving anisotropic pump- and loss-rates (see Sect.~3.3),
   the pump and loss events have been assumed to be isotropic here. Following Watson \& Wyld
   (2001), the approach of phenomenological maser theory is used. I model the
   polarization properties of the $v=1, J = 1-0$ transition with rest frequency
   $\omega_{\rm R} = 2.71\,\times 10^{11}\,{\rm s^{-1}}$, allowing for a
   straightforward analytical formulation of the problem. The rate equations
   yield the population differences $n_+$, $n_0$ and $n_-$ between the
   split-up upper state ($J=1, M = +1, 0, -1$, respectively) and the lower
   state ($J=0, M=0$).
   These population differences are normalized by the ratio of the differential 
   pump rate (into the $J=1$ respectively $J=0$ levels) to the loss rate
   (assumed to be the same for both levels), and are a function of
   molecular velocity. Because of the isotropic pumping by collisions or
   unpolarized radiation assumed here, the pump rates are equal into the sublevels
   of the $J=1$ state. Collisionally induced transitions between magnetic sublevels of a given
   $J$ state are neglected. This yields
   \begin{eqnarray} 
   \lefteqn{n_+ =} \nonumber \\
   & \frac{\scriptstyle (1+R_-)(1+R_0)}{\scriptstyle (1+2R_+)(1+R_0)(1+R_-)
                                                +R_0(1+R_-)(1+R_+)+R_-(1+R_+)(1+R_0)}, \\
   \lefteqn{ } \nonumber \\
   \lefteqn{n_0 =} \nonumber \\
   & \frac{\scriptstyle (1+R_+)(1+R_-)}{\scriptstyle (1+2R_+)(1+R_0)(1+R_-)
                                                +R_0(1+R_-)(1+R_+)+R_-(1+R_+)(1+R_0)}, \\
   \lefteqn{ } \nonumber \\
   \lefteqn{n_- =} \nonumber \\
   & \frac{\scriptstyle (1+R_+)(1+R_0)}{\scriptstyle (1+2R_+)(1+R_0)(1+R_-)
                                                +R_0(1+R_-)(1+R_+)+R_-(1+R_+)(1+R_0)}\,.
   \label{eq:population}
   \end{eqnarray}
   Here the stimulated emission rates $R_\pm$ and $R_0$ for the transitions 
   with $\Delta M = \pm 1$ and $0$ are normalized by the
   saturation intensity $I_{\rm S} = 8 \hbar\omega^3 \Gamma/3\pi c^2 A$ 
   (A being the Einstein coefficient for spontaneous emission) and
   are, for a given observing frequency $\omega$, evaluated at frequencies 
   $\omega\mathcal{D}$, where $\mathcal{D}$ are the Doppler factors given in 
   Table~\ref{table:1}.
%
\begin{table}
\caption{Doppler factors for evaluation of the stimulated emission rates}
\centering          
\begin{tabular}{c c c c }     
\hline       
normalized stimulated & \multicolumn{3}{c}{Doppler factors $\mathcal{D}$ for 
evaluating}  \\
emission rate & $n_+$ & $n_0$ & $n_-$\\
\hline                    
      &     &      &     \\
$R_+$ &  1  & \large $\frac{\omega_{\rm R}+g\Omega/2}{\omega_{\rm R}}$ &
              \large $\frac{\omega_{\rm R}+g\Omega/2}{\omega_{\rm R}-g\Omega/2}$ \\
      &     &      &     \\
$R_0$ & \large $\frac{\omega_{\rm R}+g\Omega/2}{\omega_{\rm R}}$ & 1 & 
        \large $\frac{\omega_{\rm R}}{\omega_{\rm R}-g\Omega/2}$                 \\
      &     &      &     \\
$R_-$ & \large $\frac{\omega_{\rm R}-g\Omega/2}{\omega_{\rm R}+g\Omega/2}$ &
        \large $\frac{\omega_{\rm R}-g\Omega/2}{\omega_{\rm R}}$           &  1   \\
      &     &      &     \\
\hline
\end{tabular}
\label{table:1}      
\end{table}
%

   These Doppler factors account for the fact that molecules at different
   line-of-sight velocities - due to their velocity distribution and,
   if present, a velocity gradient (not considered here) - couple to different
   Zeeman components. The stimulated emission rates $R_{\pm}$ and $R_0$ are
   given by (Goldreich et al. 1973)
   \begin{eqnarray}
   R_{\pm} & = & I_\pm(1+\cos^2{\gamma})+\sin^2{\gamma}(Q_\pm\cos{2\eta}
                 -U_\pm\sin{2\eta}) \nonumber \\
           &   & \pm 2V_\pm\cos{\gamma}\,, \\
   R_0     & = & 2 \sin^2{\gamma}(I_0-Q_0\cos{2\eta}+U_0\sin{2\eta})
   \label{eq:stimRate}
   \end{eqnarray}
   where $I,Q,U$ and $V$ are the Stokes parameters, normalized by the 
   saturation intensity and multiplied\footnote{For a linear maser
   considered here, intensity means its solid-angle average multiplied by $4\,\pi$.}
   by $4\pi$, and the indices $\pm,0$ indicate the frequency at which these
   quantities are to be evaluated (Table~\ref{table:1}). Here, $\gamma$ is the
   angle between the magnetic field and the line-of-sight, directed outwards from the
   drawing plane of Fig.\,\ref{fig:sketch}. $\eta$ is the angle between the $z$ axis
   and the projection of the magnetic field onto the plane of the sky (i.e., the drawing
   plane in Fig.\,\ref{fig:sketch}), $B_{\rm sky} = \sqrt{B_{\rm y}^2+B_{\rm z}^2}$. These
   angles are calculated from
   \begin{equation}
	\cos{\gamma} = \frac{B_{\rm x}}{|| {\bf B} ||} \,,
	\cos{\eta}   = \frac{B_{\rm z}}{|| {\bf B} ||} \,.
   \label{eq:gammaEta}
   \end{equation}
   For a static magnetic field, the coordinate system is conveniently chosen
   such that $\eta=0^\circ$ (i.e., the system in which Stokes U vanishes in
   the absence of magnetorotation).

\subsection{Radiative transfer}

   Since the light-travel time is negligible with respect to the time scale
   on which the magnetic field will vary, the stationary radiative
   transfer equation is used:
   \begin{eqnarray}
   \frac{d}{d\tau} \left( \begin{array}{c} I \\
                                           Q \\
					   U \\
					   V
                          \end{array}
                   \right) & = & 
			  \mathbf{K} \left(
			                   \begin{array}{c} I \\
			                                    Q \\
				                            U \\
				                            V
				            \end{array}
				     \right)         \,.
   \label{eq:radtrans}
   \end{eqnarray}
   As usual in maser theory, the source term taking spontaneous emission
   into account is insignificant here, and $\tau$ is the opacity
   along the line of sight for the polarized plus unpolarized radiation.
   The M\"uller matrix $\mathbf{K}$ of the absorption coefficients has the
   elements
   \begin{eqnarray}
   \mathbf{K} & = & \left( \begin{array}{cccc} 
                             A & B    & F    & C \\
			     B & A    & E    & G \\
			     F & $-E$ & A    & D \\
			     C & $-G$ & $-D$ & A
                           \end{array}
                    \right)\,.
   \label{eq:MuellerMatrix}
   \end{eqnarray}
   If the vertical polarization is along the projection of the magnetic field onto the plane
   of the sky, $\mathbf{B_{\rm sky}}$, the elements of $\mathbf{K}$ have the following meaning:
   Coefficients $A,B,C$ and $F$ are the absorption coefficients for different
   polarization states, while $D, E$ and $G$ describe anomalous dispersion
   effects (Landi degl'Innocenti \& Landi degl'Innocenti 1981). Elements
   $F$ and $G$ are assumed to be negligible, since $g\Omega \gg R$. This 
   condition is necessary here, otherwise nothing about the magnetic field
   can be inferred from measurements of $p_{\rm C}$, since a given
   molecular state would be deexcited by a stimulated emission before a full
   Larmor precession is accomplished. The matrix elements $A, B,$ and $C$ are,
   for a two-level system with split-up upper level, given by (Watson \& Wyld
   2001),
   \begin{eqnarray}
   A & = & (1+\cos^2{\gamma})(f_+^{\rm (r)}n_+ +f_-^{\rm (r)}n_-)
           +2f_0^{\rm(r)} n_0\sin^2{\gamma}\,, \\
   B & = & \sin^2{\gamma}(f_+^{\rm (r)}n_+ +f_-^{\rm (r)}n_-
           -2f_0^{\rm (r)}) \,, \\
   C & = & 2 \cos{\gamma}(f_+^{\rm (r)}n_+ -f_-^{\rm (r)}n_-)
           +2f^{\rm(r)}_0 n_0\sin^2{\gamma}\,.
   \label{eq:abc}
   \end{eqnarray}
   The terms $D$ and $E$ describe magneto-rotation, i.e., the conversion of 
   linear to circular polarization, or -in other words - the generation of the
   latter by means other than the Zeeman effect. They are given by 
   \begin{eqnarray}
   D & = & \sin^2{\gamma}(f_+^{\rm (i)}n_+ +f_-^{\rm (i)}n_-
           -2f_0^{\rm (i)}) \,, \\
   E & = & 2 \cos{\gamma}(f_+^{\rm (i)}n_+-f_-^{\rm (i)}n_-)
           +2 \frac{d\eta}{d\tau}
   \label{eq:de}
   \end{eqnarray}
   where the extra term $2d\eta/d\tau$ in $E$ accounts for the rotation of the
   coordinate system, while the photon propagates along the line-of-sight, such
   that vertical polarization is always along $B_{\rm sky}$. This extra term
   is a lengthy expression calculated from the magnetic field Eqs.~19 to 21
   and the transformation defined by Eqs.~\ref{eq:Euler1},\,\ref{eq:Euler2}
   (see below). To correctly handle phase shifts induced by magneto-rotation,
   the profile function $f$ is complex, where $f_+^{\rm (r)}$ and
   $f_+^{\rm (i)}$ are its real and imaginary part, respectively, at the
   normalized frequency offset
   \begin{equation}
   \upsilon_+ = \left(\omega-[\omega_0+g\Omega/2]\right)
    \frac{1}{\Delta\omega_{\rm D}}
   \label{eq:upsilon+}
   \end{equation}
   where $\Delta\omega_{\rm D}$ is the Doppler width of the line, and $\omega_0$
   the resonance rest frequency of the transition. Here I use
   $\Delta\omega_{\rm D} = 1$\,km\,s$^{-1}$, which corresponds to the thermal linewidth
   of SiO in a gas at 1000\,K.

   Correspondingly, $f_-^{\rm (r)}$ and $f_-^{\rm (i)}$ are the real and imaginary parts
   of the profile function at 
   \begin{equation}
   \upsilon_- = \left(\omega-[\omega_0-g\Omega/2]\right)
                \frac{1}{\Delta\omega_{\rm D}}
   \label{eq:upsilon-}
   \end{equation}
   and $f_0^{\rm (r)}$ and $f_0^{\rm (i)}$ at 
   \begin{equation}
   \upsilon_0 = \left(\omega-\omega_0\right) \frac{1}{\Delta \omega_{\rm D}}\,.
   \label{eq:upsilon0}
   \end{equation}
   The complex profile function $f(\omega)$ is given by 
   \begin{equation}
   f(\omega) =  
   \frac{1}{\pi} \int_{-\infty}^{+\infty}{
                  \frac{\Phi(v)dv}{\Gamma+i\left(
                                       \omega_0-\omega[1-\frac{v}{c}] \right) }}
   \label{eq:profile}
   \end{equation}
   where $\Gamma$ is the decay rate of the excited state ($5\,\rm{s}^{-1}$ for
   the SiO masers considered here, see Nedoluha \& Watson 1994), and
   $\Phi(v)$ is the velocity distribution of the masing gas, which is taken to be
   Maxwellian. The velocity distribution of molecules in different substates differs
   from a pure Maxwellian distribution, because Eqs.~1-3 are velocity-dependent.
   The real and imaginary parts of $f(\omega)$ are known as the Voigt function
   $\mathcal{H}$ and the Faraday-Voigt function $\mathcal{F}$, respectively, namely
   \begin{equation}
   f(\upsilon,a) = \frac{1}{\sqrt{\pi}}\left[\mathcal{H}(a,\upsilon)
   +2i\,\mathcal{F}(a,\upsilon)\right]
   \label{eq:VoigtFaraday}
   \end{equation}
   with $a = \Gamma / \Delta\omega_{\rm D}$. For $\mathcal{H}$ and $\mathcal{F}$
   I used the rational approximation given by Humlicek (1982), as provided in
   an optimized form by Schreier (1992). 

   \begin{figure*}[ht!]
   \centering
   \includegraphics[width=12cm]{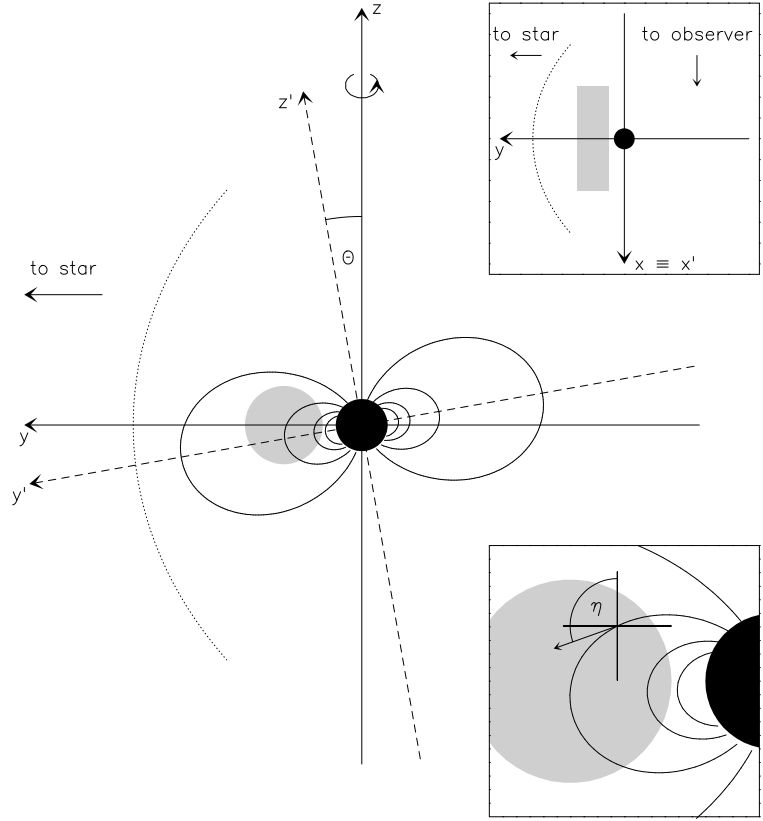}
   \caption{Sketch of the model with magnetic field lines. The $x,y,z$
            coordinate system (solid lines) is defined by the rotation axis
            of the planet ($z$ axis) and the line of sight to the observer
            ($x$ axis, directed from the drawing plane outwards). The
            coordinate system $x',y',z'$ (dashed lines) of the magnetosphere
            is defined by the tilt angle $\Theta$ of the dipole axis with
            respect to the rotation axis. The rotation phase here is chosen
            such that the $x'$ axis is identical to the $x$ axis. The cross
            section of the planet is depicted by the black circle, the maser
            slab by the grey one (for models 1b, 2b). The location of the
            Chapman-Ferraro magnetopause is indicated by the dotted line.
            The top right insert shows the $(x,y)$ plane with the
            location of the magnetopause (dashed line) and the maser slab as
            a grey rectangle (field lines omitted). The bottom right insert shows
            a zoom onto a particular line of sight, marked by a cross, within
            the SiO maser bundle. The arrow indicates the local magnetic field
            direction and the definition of angle $\eta$.}
         \label{fig:sketch}
   \end{figure*}
%
   Without a detailed model at hand, I include neither the
   magnetopause nor the magnetotail, but assume that the SiO maser slab crosses 
   a magnetic dipole field (in spherical coordinates, defined by the symmetry
   axis of the dipole field, e.g., Lewis 2004), given by
   \begin{eqnarray}
	B_{\rm r} & = & \frac{2M}{r^3} \cos{\theta}, \\
	B_\theta  & = & \frac{M}{r^3}  \sin{\theta}, \\
	B_\phi    & = & 0\,{\rm G},
   \label{eq:Bfield}
   \end{eqnarray}
   where $M = 34 {\rm G} r_{\rm J}^3$ is the magnetic
   dipole moment ($r_{\rm J}$ is Jupiter's mean radius). The justification
   of using such a magnetic field component is discussed in Appendix\,A.
   We now transform the magnetic dipole field from its stationary frame
   into a coordinate system defined by the planetary rotation axis,
   ${\bf e}_{\rm z} = (0,0,1)^{\rm T}$, the line-of-sight towards the
   observer, ${\bf e}_{\rm x} = (1,0,0)^{\rm T}$ and
   ${\bf e}_{\rm y} = (0,1,0)^{\rm T}$, such that
   $({\bf e}_{\rm x},{\bf e}_{\rm y},{\bf e}_{\rm z})$ form a
   left-handed coordinate system (Fig.\,\ref{fig:sketch}). The transformation
   is defined by the Euler angles $\Phi = 2\pi t/T = \Omega t$ (where $T$ is
   the rotation period, and $\Theta$ the inclination of the
   magnetic dipole axis against the rotation axis), and is obtained by
   \begin{eqnarray}
	    \left(
	    \begin{array}{c}         
		 x'\\
		 y'\\
		 z'\\
	    \end{array} 
	    \right)
	    & = & M(t) \left(
		       \begin{array}{c}
			    x \\
			    y \\
			    z \\
		       \end{array}
		       \right)
   \label{eq:Euler1}
   \end{eqnarray} 
   with
   \begin{eqnarray}
	    M(t) & = & 
		       \left(
			     \begin{array}{rrr}

	    \cos{\Omega t}             &  \sin{\Omega t}             &
		       0 \\
	   -\cos{\Theta}\sin{\Omega t} &  \cos{\Theta}\cos{\Omega t} &
	    \sin{\Theta} \\
	    \sin{\Theta}\sin{\Omega t} & -\sin{\Theta}\cos{\Omega t} &
	    \cos{\Theta}
			      \end{array}
		       \right)
   \label{eq:Euler2}
   \end{eqnarray}
   (in Fig.\,\ref{fig:sketch}, $\Phi = 90^\circ$ has been chosen).
   For $T$ and $\Theta$, values close to those for Jupiter are used,
   i.e., $T = 10$\,h and $\Theta = 10^\circ$, assuming that the planet
   is still too far away from the star to lose angular momentum to spin up
   the latter (Soker 2001). Keeping the typical values for giant gas
   planets in the solar system is therefore not too farfetched in this context.
%
\begin{table*}
\caption{Summary of models}             
\centering          
\begin{tabular}{l r r r r r r}
\hline       
\hline       
resonance frequency            & $\omega_{\rm R} $      & $=$ & \multicolumn{4}{l}{$2.71\,\times 10^{11}$\,s$^{-1}$} \\
FWHM  of velocity distribution & $\Delta v$             & $=$ & \multicolumn{4}{l}{1.67\,km\,s$^{-1}$} \\
planetary magnetic moment      & $M $                   & $=$ & \multicolumn{4}{l}{34\,G\,$r_{\rm J}^3$}\\
tilt of magnetic dipole axis   & $\Theta $              & $=$ & \multicolumn{4}{l}{$10^\circ$} \\
maser gain length              & $L$                    & $=$ & \multicolumn{4}{l}{10\,$r_{\rm J}$} \\
\hline  
Model                          & 1a & 1b & 2a & 2b & 2c & 2d \\
\hline
distance of maser slab from rotation axis [$r_{\rm J}$] & 3  & 3  & 3   & 3   & 0   & 0.5 \\
height above equatorial plane  [$r_{\rm J}$] & 0  & 0  & 0   & 0   & 3   & 3   \\
diameter of maser bundle       [$r_{\rm J}$] & 0  & 3  & 0   & 3   & 0   & 0   \\
normalized peak intensity                    & 10 & 10 & 0.1 & 0.1 & 0.1 & 0.1 \\
\hline  
\hline  
\end{tabular}
\label{table:2}      
\end{table*}
%
\section{Results and discussion}

In the following, the model results (Figs.\,\ref{fig:saturatedPencil}
to \ref{fig:unsaturatedPencil3}) are shown as grey-scale plots
(fractional polarization or polarization angle vs. phase and velocity) and
as polarization spectra at a given reference phase of the planetary rotation,
or as a time series at a given reference velocity. To search for strong
magnetic fields, maser features with a strong fractional circular polarization
($p_{\rm C}$) are needed (which also suffer less than linear polarization from
the confusion in the main beam of the spatially unresolved observations).
I therefore use the phase and velocity of the maser feature with
the strongest $p_{\rm C}$ as reference. 
\subsection{Saturated maser}
%
   \begin{figure}[ht!]
   \includegraphics[width=6.5cm,angle=-90]{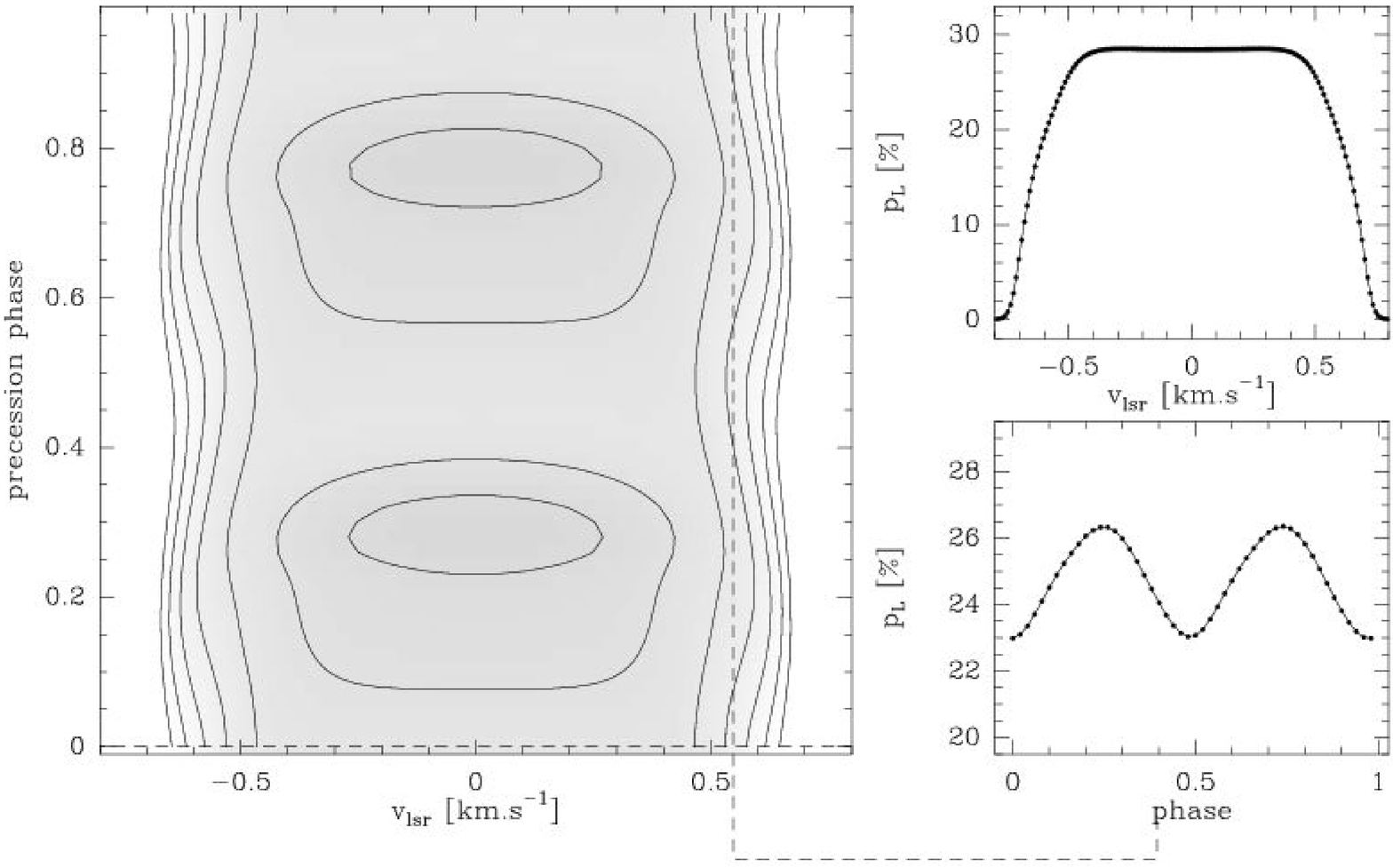}
   \includegraphics[width=6.5cm,angle=-90]{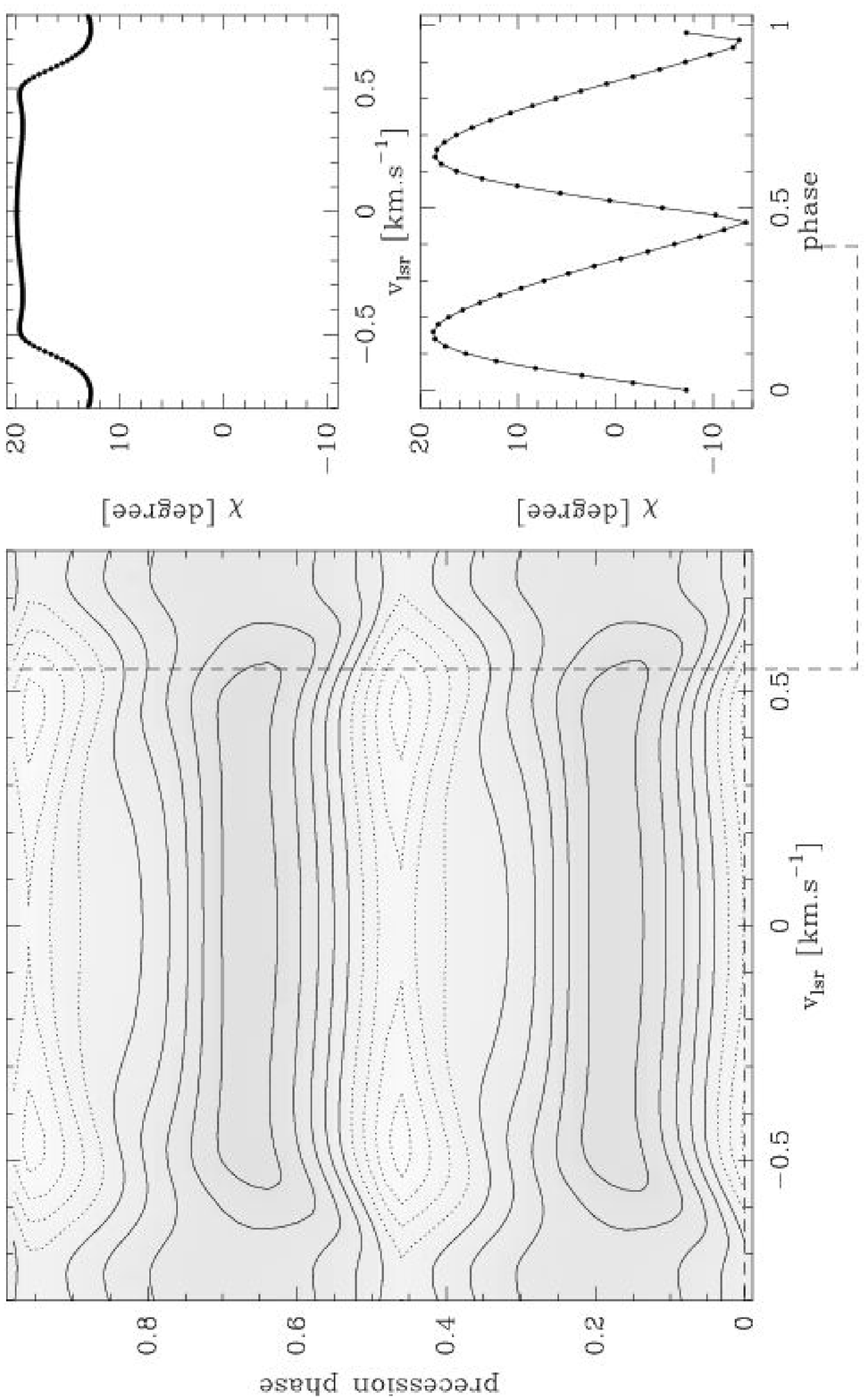}
   \includegraphics[width=6.5cm,angle=-90]{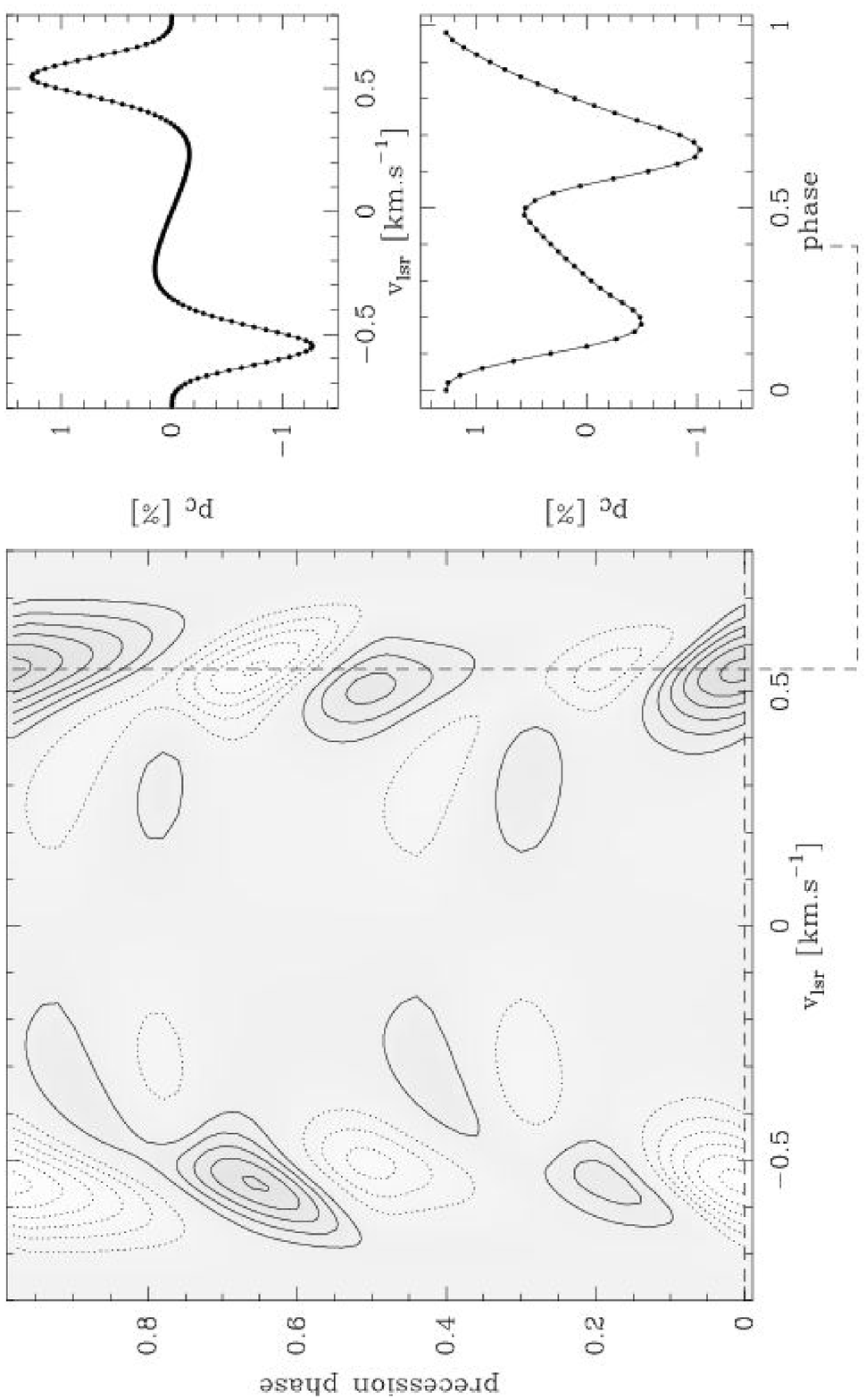}
   \caption{Saturated SiO maser ($I_{\rm peak} = 10\,I_{\rm S}$),
            pencil beam at $r=3r_{\rm J}$, in the planet's equatorial
            plane (model 1a). Inserts: spectra and time series at
            the phase (indicated by a horizontal dashed line),
            respectively velocity (indicated by a vertical dashed
            line), of maximum circular polarization. Each dot
            represents a point of the time sampling.
            Top: fractional linear polarization ($p_{\rm L}$ vs.
            velocity and phase (contour levels $15\,\%$ to $33\,\%$ by 
            $3\,\%$). Center: position angle $\chi$ of linear
            polarization East from North, contour levels $-18^\circ$
            to $+18^\circ$ by $4^\circ$). Bottom: fractional circular
            polarization ($p_{\rm C}$, contour levels from $-1.2\,\%$
            to $+1,2\,\%$ by $0.2\,\%$, zero contour suppressed).
            \label{fig:saturatedPencil}
           }
   \end{figure}
The first model 1a (see Table\,2 and Fig.\,\ref{fig:saturatedPencil}) is for
a maser with a pencil beam in the planet's equatorial plane, at $3\,r_{\rm J}$
from the center of the planet, with a length of $10\,r_{\rm J}$. The opacity is
scaled such that the peak intensity is $10\,I_{\rm S}$, i.e., the maser is saturated.
Because of both the small tilt of the magnetic dipole axis against the rotation axis, and
the absence of a toroidal magnetic field component $B_\phi$, $\cos{\gamma} \simeq 1$ and
the linear polarization does only slightly exceed the canonical value of 33\,\%
(Goldreich et al. 1973) for a saturated maser propagating perpendicular to the
magnetic field. (Our excess stems from line-of-sight elements with $\cos^2{\gamma} > 2/3$.) 
The peak-to-peak variation in the fractional linear polarization, $p_{\rm L}$, 
is 6.6\,\% at line center, and 3.4\,\% at the velocity of maximum circular polarization.
The circular polarization displays the archetypical S-shaped line profile of an
unresolved Zeeman pattern whose peak-to-peak amplitude varies in time by $2.6\,\%$.
This effect will be observable provided that the maser is sufficiently strong.
The relatively strong circular polarization is not only produced by the Zeeman
effect (term C in Eq.~\ref{eq:MuellerMatrix}), but also increased by the rotation of the
magnetic field in the maser amplification zone along the line-of-sight, rotating the
linear polarization from Stokes Q to Stokes U via the E-term in
Eq.~\ref{eq:MuellerMatrix}, and then its conversion from Stokes U to Stokes V via the
D-term.

In observational practice, the situation is often less clear, for two reasons. First,
while polarization monitoring with single-dish telescopes offers the advantage of being free
of fluctuations due to baseline synthesis effects in VLBI observations, it lacks the spatial
resolution of the latter. What is obtained is instead a spectral blend of Zeeman features from
individual maser spots, centered at different velocities. If the full width to zero Stokes V
from an individual feature is below the velocity separation of two spatially separated 
maser features, observations with a high enough spectral resolution will help. If not, this
results in confusion. Even the spectra of spatially resolved maser spots may be concerned in the
case of a velocity gradient along a single maser slab, leading to velocity redistribution. As already stated
above, the latter case is not handled here, with the rationale of restricting the models to the basic
cases before dealing with the complications (which, in the limit of a uni-directional slab maser,
can of course be handled by an extension of the theory described above). Nature may help us here, because the
magnetic flux density in a planetary magnetosphere may be enhanced with respect to that of a magnetic
field of stellar origin, but this hope needs to be ascertained. We cannot expect that all detections of a
strong Stokes V in SiO masers (corresponding to a few up to 20~G, Herpin et al. 2006) are owing to
planetary magnetospheres.

The second caveat concerns both VLBI and single-dish observations.
In reality, SiO masers in Mira stars have a typical size of at least
$\sim 10^{12}$\,cm (constrained both from VLBI observations, e.g., Philips \& Boboltz
2000, and models, Bujarrabal 1994), which is much larger than the length scale on which
the magnetic dipole field varies in model 1a. To simulate the effects of
diluting the magnetosphere in the maser cross section, a cylindrical maser slab
(model variant 1b) has also been modeled as a collection of maser rays with a bundle
diameter of $3\,r_{\rm J}$ (i.e., two orders of magnitude below observed spot sizes). This
corresponds, at a distance of e.g., 230~pc, to an angular size of 0.01~mas (i.e.,
still unresolved by VLBI observations). Assuming a larger and more realistic size would
require including the magnetopause and the magnetotail in the model, and possibly the detailed
dynamics of mass accretion onto the planet, which clearly is beyond the scope of this
paper. I therefore assume that the polarization signals modeled here are on top
of a linear or circular polarization background, which does not vary on a timescale
comparable to that of the planetary rotation. We know that the lifetime of SiO masers is
unlikely to be longer than a cycle of the stellar variability, because the features are disrupted by 
shock waves driven by the stellar pulsation (Humphreys et al. 2002). Multiepoch VLBA imaging of
SiO $v=1, J=1-0$ masers impressively confirmed this picture (the ``movie'' of TX\,Cam, Diamond \& Kemball,
2003). Long-term polarization fluctuations may occur on shorter timescales (several
weeks at least, Glenn et al. 2003). They come from slow readjustments of the magnetic field. These
timescales are long enough to monitor at least 10 periods of the planetary rotation, allowing us to 
separate the short-term polarization fluctuations modeled here from the long-term ones of polarized
and total maser flux.

\begin{figure}[h!]
   \includegraphics[width=6.5cm,angle=-90]{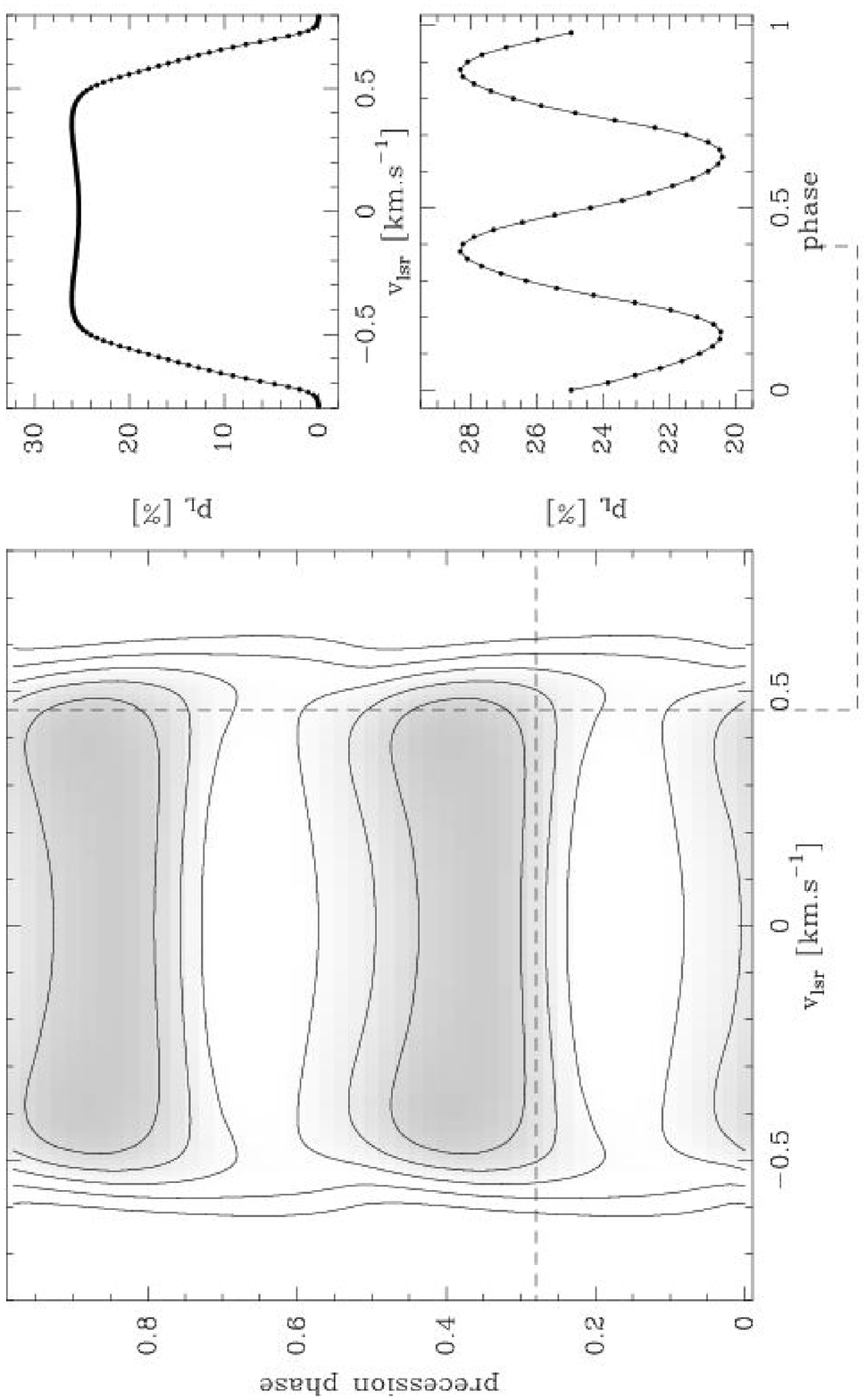}
   \includegraphics[width=6.5cm,angle=-90]{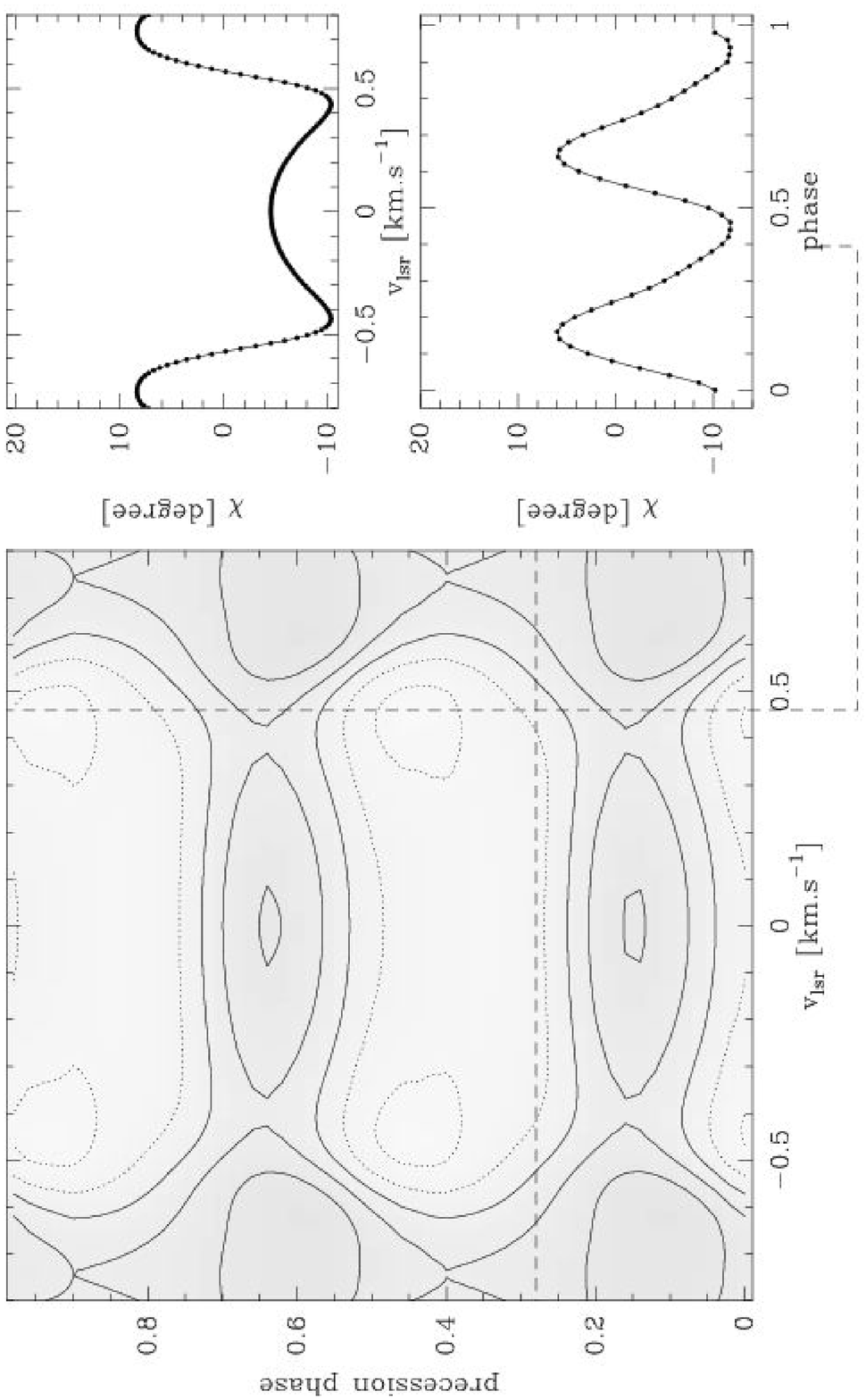}
   \includegraphics[width=6.5cm,angle=-90]{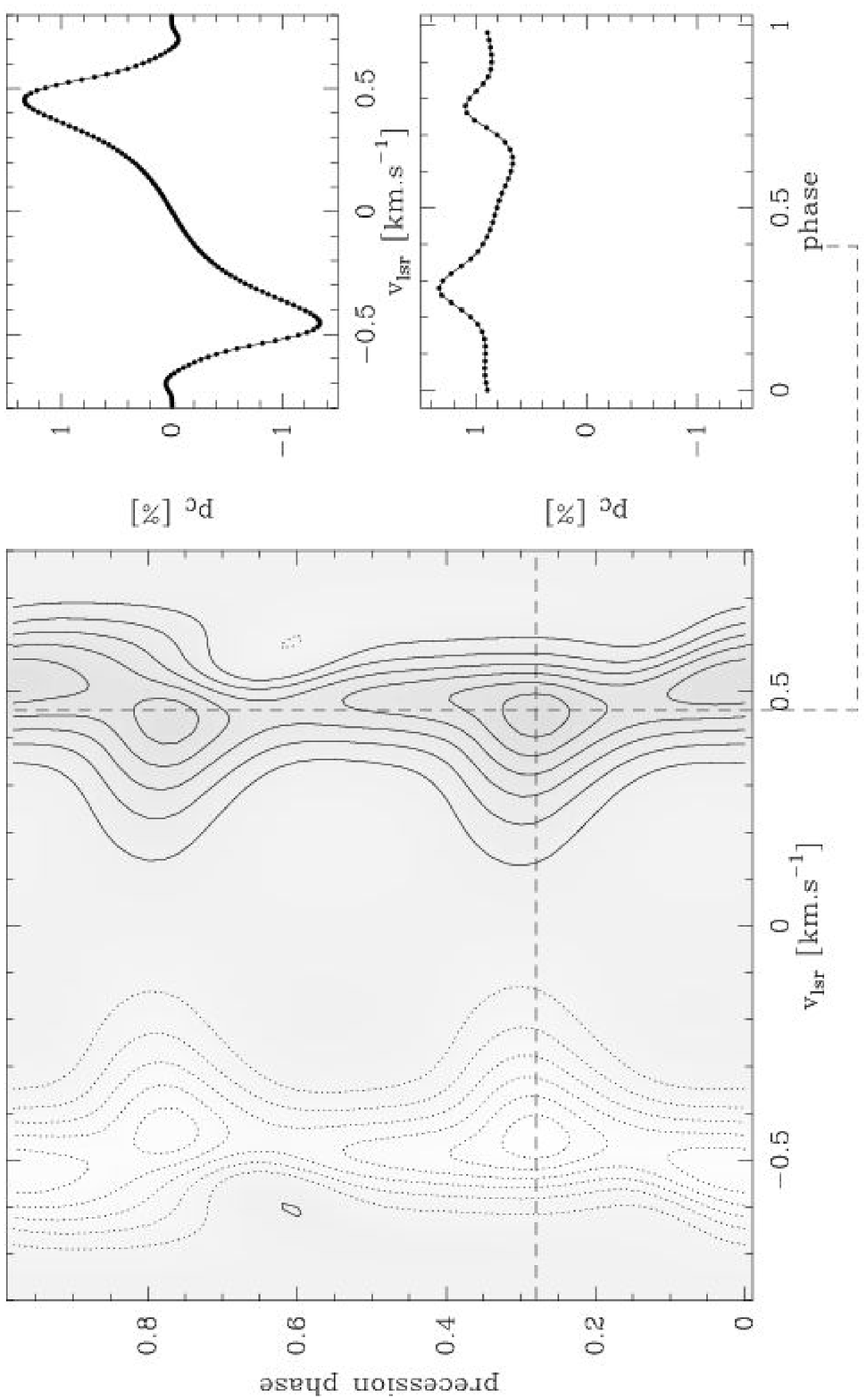}
   \caption{Same as Fig.~\ref{fig:saturatedPencil}, but for a bundle
            of SiO masers, with a slab diameter of $3\,r_{\rm J}$
            (model 1b). For easy comparison, the plot limits for the
            inserts are the same as those for the equivalent pencil
            beam model (Fig.~\ref{fig:saturatedPencil}). Top:
            $p_{\rm L}$, contour levels from $15\,\%$ to $27\,\%$ by
            $3\,\%$. Center: $\chi$, contour levels from $-10^\circ$
            to $+10^\circ$ by $5^\circ$. Bottom: $p_{\rm C}$, contour
            levels from $-1.2\,\%$ to $+1.2\,\%$ by $0.2\,\%$ (zero
            contour omitted).
            \label{fig:saturated}
           }
   \end{figure}

The polarization background is not necessarily dominant, even in single-dish observations. Linear
polarization is reduced by the combined action of both the predominantly tangential polarization
of the maser spots distributed along a ring (providing indirect evidence of radiative pumping,
Desmurs et al. 1999) and cancellation of polarization vectors in the observing beam. As for circular
polarization, we may have reasons to expect that it is enhanced towards the lines of sight crossing the
strongest magnetic fields, i.e., close to the planet, as suggested above. Any observational attempt
to discover the features modeled here will therefore focus on measurements of the circular
polarization. Fig.\,\ref{fig:saturated} shows that the maximum linear polarization is
reduced by about $5.3\,\%$ due to dilution (even in VLBI observations, since they do 
not resolve the maser spot either), and incoherent mixing correspondingly reduces the
peak-to-peak variation of the polarization angle. As expected, the circular polarization
basically remains the same. (Because of the mixing of maser rays with smaller Zeeman
splittings, the velocity of maximum $p_{\rm C}$ is reduced.) At the velocities of maximum 
circular polarization, the relative peak-to-peak variation of the linear polarization has 
doubled with respect to the pencil-beam model (1a), thanks to the shift in the velocity of
maximum circular polarization (which is used as reference).
\subsection{Unsaturated maser}
\begin{figure}[h!]
   \includegraphics[width=6.5cm,angle=-90]{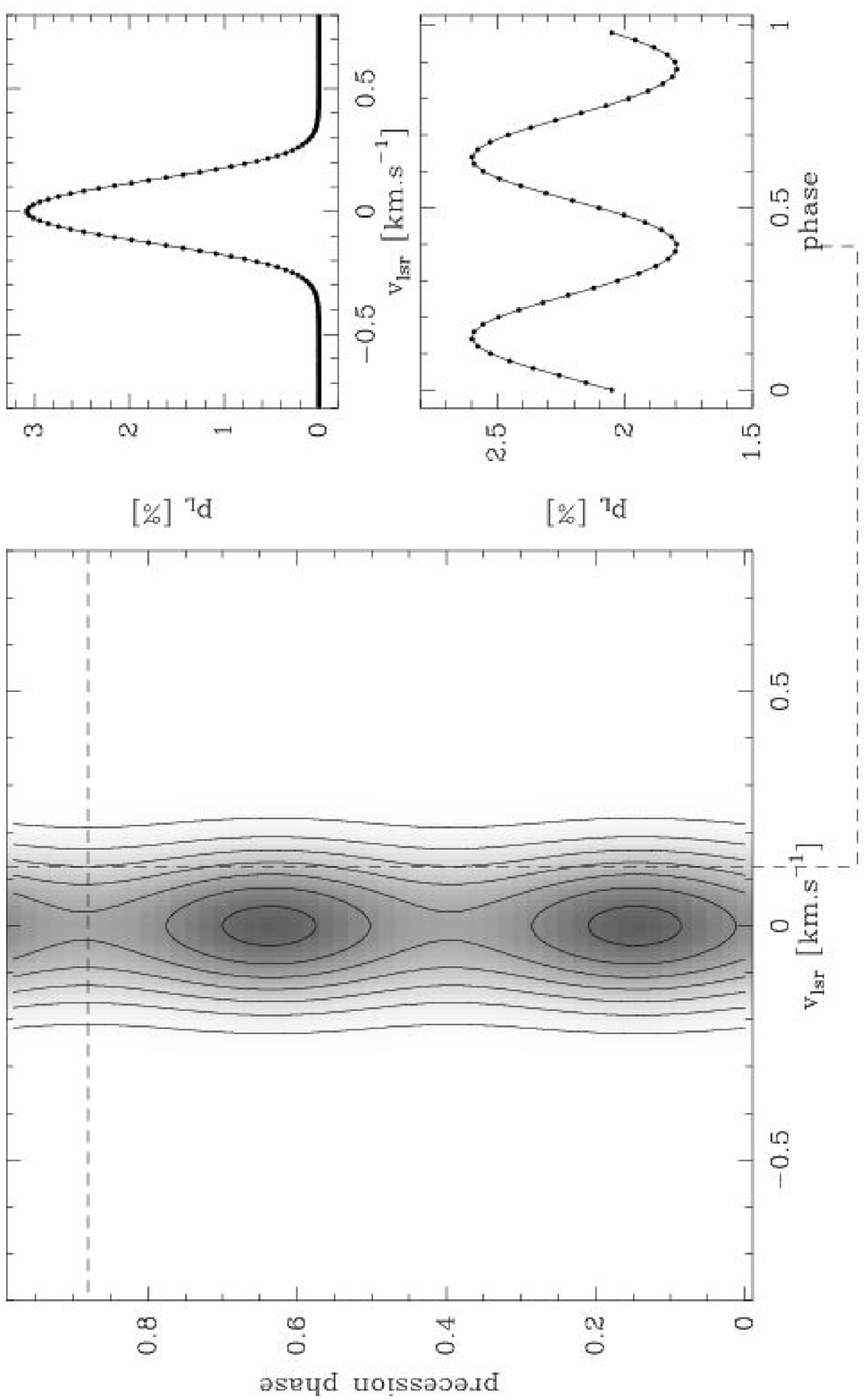}
   \includegraphics[width=6.5cm,angle=-90]{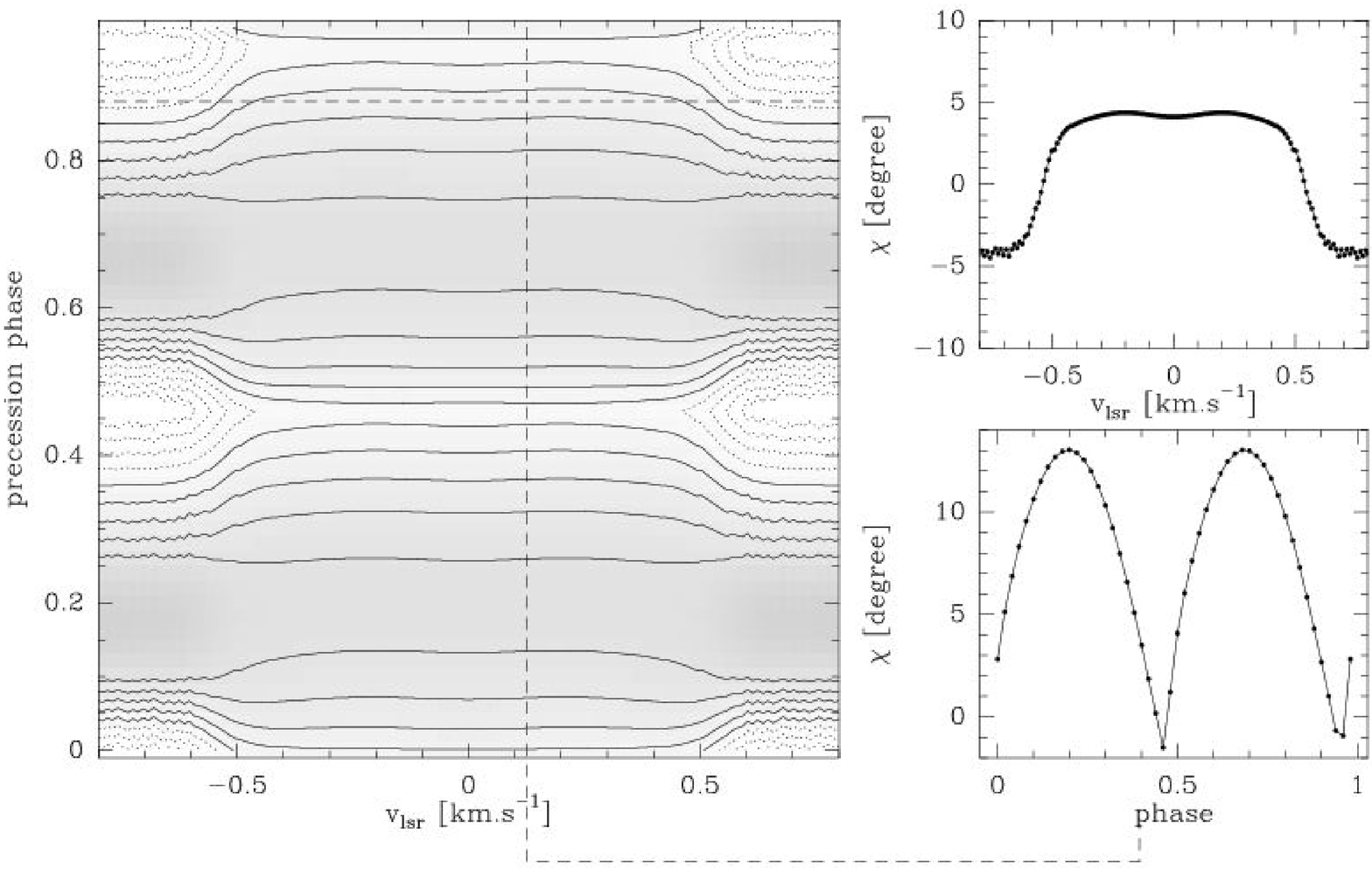}
   \includegraphics[width=6.5cm,angle=-90]{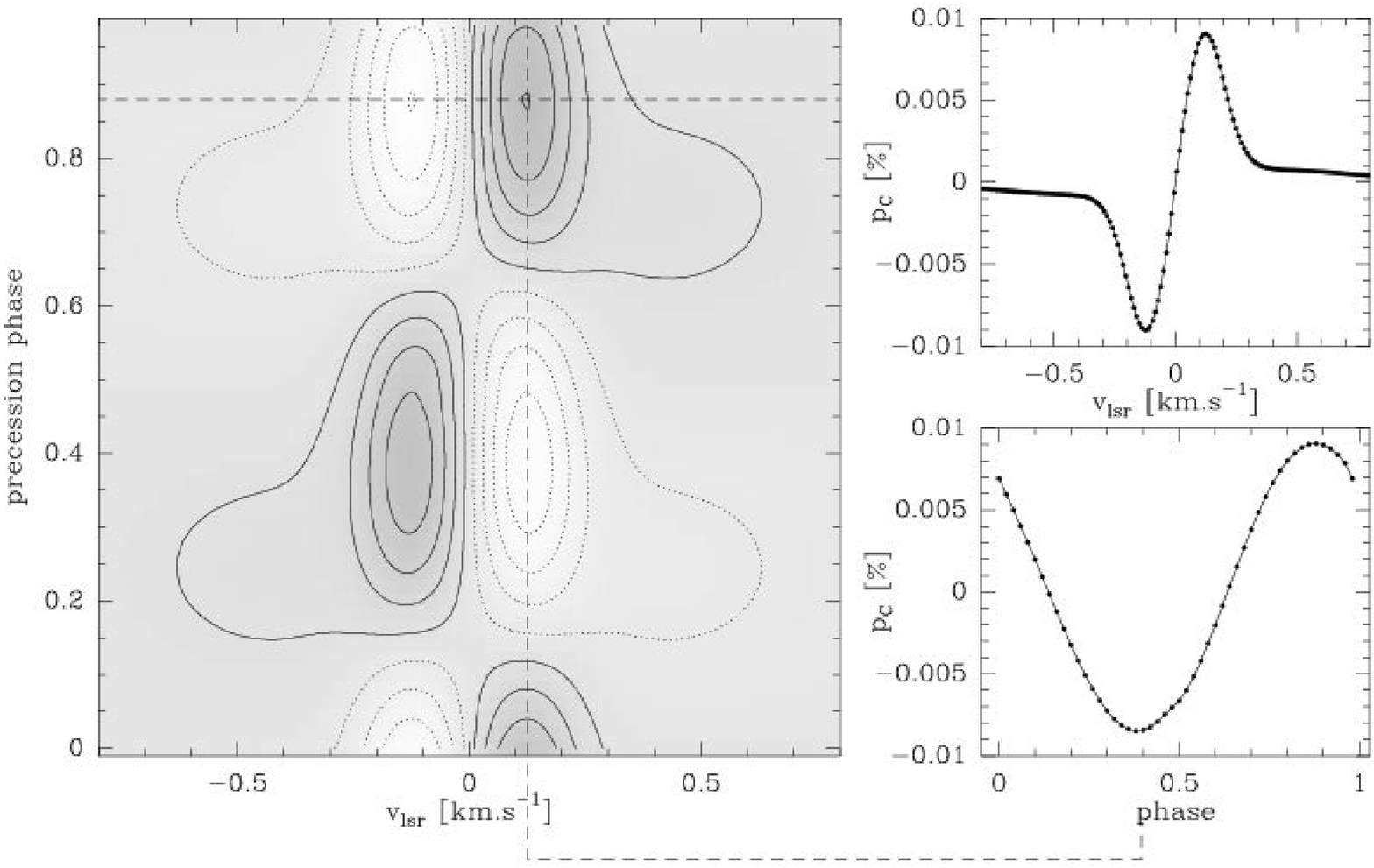}
   \caption{Unsaturated SiO maser ($I_{\rm peak} = 0.1\,I_{\rm S}$),
            pencil beam at $r=3\,r_{\rm J}$, in the planet's
            equatorial plane (model 2a). The linewidths are narrowed
            with respect to the thermal width, as expected for
            unsaturated maser action. Top: $p_{\rm L}$, contour levels 
            from $0.6\,\%$ to $4.2\,\%$ by $0.6\,\%$. Center: $\chi$,
            contour levels $-12^\circ$ to $12^\circ$ by $3^\circ$.
            Bottom: $p_{\rm C}$, contour levels from $-0.009\,\%$ to
            $+0.009\,\%$ by $0.002\,\%$.
            \label{fig:unsaturatedPencil}
           }
   \end{figure}

The saturation of masers is difficult to determine experimentally, mainly due to its 
dependence on the observationally poorly constrained beaming angle (Watson \& Wyld 2001).
The polarization of unsaturated masers is weaker than that of their saturated counterparts,
because only the latter convert each pumping event into a maser photon.
A fully unsaturated maser ($\Gamma \ll R$) is unpolarized if both the seed and pump radiation are 
unpolarized. The model 2a (Table\,2 and Fig.\,\ref{fig:unsaturatedPencil}) is for
a pencil beam with $0.1\,I_{\rm S}$, and yields a maximum linear polarization of 4.4\,\%
at line center (with a peak-to-peak variation of 1.3\,\%), which drops to 2.6\,\% at the
velocity of strongest circular polarization (with a peak-to-peak amplitude of $0.8\,\%$). 
As expected in unsaturated masers, the line profile of Stokes
V is identical to that of thermal radiation (Fiebig \& G\"usten 1989, see also Watson
\& Wyld 2001), and the linewidths are narrowed with respect to the thermal width
(cf. Watson \& Wyld, Fig.\,2). Therefore, the circular polarization is
drastically reduced (to $p_{\rm C} = \pm 0.009\,\%$) for a maser ray in the equatorial
plane, and is no longer detectable with current sensitivity. Again, a cylindrical maser
slab (model 2b, Fig.\,\ref{fig:unsaturated}, electronic version) does not drastically
alter the result for circular
polarization, but reduces the linear polarization at the reference channel to a maximum
of 1.5\,\%, with a peak-to-peak variation of 0.3\,\% (i.e., mean value and amplitude are
scaled down by a factor $1:2$).

For the detectability of the circular polarization of an unsaturated SiO maser from a
Jovian magnetosphere, its location in the latter is important. Because the circular
polarization here is proportional to the line-of-sight component of the magnetic field
(hereafter $B_{\rm los})$, it is naturally enhanced for maser rays crossing the rotation
axis, which have a maximum $B_{\rm los}$ before and after the crossing. This is
demonstrated by model (2c) in Fig.~\ref{fig:unsaturatedPencil2} (electronic version),
where the maser ray is otherwise at the same distance from the planet as in model (2a).
The peak-to-peak variation of $p_{\rm C} = \pm 0.29\,\%$ now becomes detectable with a
carefully tuned polarimeter (whereas the maximum linear polarization drops to
$p_{\rm L} = 1.5\,\%$).

For lines of sight close to the rotation axis, $B_{\rm sky}$ changes strongly as a
function of both velocity and rotation phase. Model 2d (Fig.\,\ref{fig:unsaturatedPencil3},
available online) is for a pencil beam $3\,r_{\rm J}$ above the equatorial plane,
and slightly offset ($0.5\,r_{\rm J}$) from the rotation axis. The dependency of the
polarization angle on both $B_{\rm los}$ and the magnetic field component projected onto
the plane of the sky, $B_{\rm sky}$ leads to a complicated pattern. The polarization angle
flips by $90^\circ$ are because of the critical $\cos^2{\gamma} = 1/3$, which decides whether the linear
polarization is along $B_{\rm sky}$ or perpendicular to it (Goldreich et al. 1973), and
of the variable maser saturation across the line profile (hence the strong spectral
variations). Observationally, these flips have been proven by VLBI polarization maps 
(i.e., Kemball \& Diamond 1997) and hint at variations of the magnetic field either due
to a planetary magnetic field as modeled here or a magnetic field component of
stellar origin.
%

\subsection{The impact of refinements of the radiative model}
\subsubsection{Anisotropic pumping}
SiO masers are believed to be radiatively pumped by infrared excitation of vibrationally
excited states (Gray et al. 2009, further references therein). The anisotropic distribution
of the pump radiation leads to an unequal population of the magnetic sublevels and therefore
enhances the linear polarization (Elitzur 1996), which generates a circular polarization in
excess of its value for isotropic pump rates if magnetorotation is present (as in the models
presented here). This birefringent conversion from linear to circular polarization would make 
a detection of fluctuations of $p_{\rm C}$ possible even in the unsaturated regime. The same
presumably holds for anisotropic loss rates. In the case of anisotropic pumping, the critical value
of $\gamma$ for producing linear polarization parallel or perpendicular to $B_{\rm sky}$
will not be the $55^\circ$ of Goldreich et al., though the linear polarization will still have
one of these orientations with respect to the magnetic field. 
\subsubsection{More complex maser geometries}
A refined model would imply not only a more adequate description of the
magnetosphere crossed by the SiO maser, but also a maser geometry more complicated than
the uni-directional cylindrical maser slabs considered here. The first step to a more
realistic description of the maser action is to consider bi-directional masers (i.e., two
opposed propagation directions in a linear maser). For a constant magnetic field,
Watson \& Wyld (2001) show that the circular polarization remains nearly unaltered
between uni- and bi-directional masers, whereas the linear polarization has a tendency
to be increased in the last (while qualitatively keeping its dependency on maser
saturation and $\cos{\gamma}$).
\subsubsection{Higher rotational quantum numbers}
The comparison between the $v=1, J=2-1$ maser transition and its $v=1,J=1-0$ counterpart
(both taken to be uni-directional) shows that their circular polarization is not drastically
different. As for linear polarization, Deguchi \& Watson (1990) found that the polarization
of the $J=2-1$ transition is actually suppressed with respect to the $1-0$ transition when
$(g\Omega)^2/\Gamma \gg R \gg g\Omega$, while observationally the former 
tends to be enhanced (McIntosh \& Predmore 1993), such that strong linear polarizations can
only be achieved by other than magnetic means (e.g., the aforementioned anisotropic
pumping, as proposed, in this context, by Watson \& Wyld 2001). However, this regime does
not allow us to analyze the magnetic field geometry by means of polarization measurements,
because a single Larmor precession of the molecule is destroyed by a stimulated emission before
its completion. I therefore did not consider this case here.
\subsection{The impact of instrumental limitations}
\subsubsection{Spatial resolution}
While the quasi-stationary linear polarization tends to decrease due to mutual cancellation within the 
beam of spatially unresolved observations, the variations in the circular polarization from
different maser spots would need to fluctuate in anticorrelation in order to cancel. This is highly
unlikely, even if all variations were caused by the effect discussed here, and even more if other
phenomena - e.g., jets or coronal flux loops, occurring on different timescales - were involved.
These scenarios are discussed in Wiesemeyer et al. (2009). The observational results of that work,
and subsequent polarization monitoring from December 2008, indicate that fluctuations of the linear
and circular polarizations may appear simultaneously (as shown here) or separately (in contrast to
the models presented in this paper). The possibility of a blend of several unrelated phenomena calls
for a coordinated observational effort as proposed in our 2008 paper. The direct
imaging of planetary wake flows in the atmospheres of the nearest AGB stars with ALMA will definitely be 
an asset. According to Cherchneff (2006), we can expect suitable molecular high-density tracers to be
sufficiently abundant in the atmosphere, thanks to the non-equilibrium chemistry dominated by shock
propagation.
\subsubsection{Spectral resolution}
The Stokes spectra shown in Figs.~2 to 7 are calculated with a frequency resolution of $2\pi\Delta\nu = 10$~kHz,
corresponding to a velocity resolution of 0.01~km\,s$^{-1}$, which in practice cannot be achieved
with existing correlation polarimeters. After smoothing with a box-like kernel, 12 frequency steps wide, the
circular peak polarization of model 1a (Fig.~2) drops from $\pm 1.27$\,\% to $\pm 0.94$\%, which is still
observable (i.e., the spectral smoothing does not lead to an order-of-magnitude loss of circular polarization).
The decrease in circular polarization due to the limited spectral resolution depends on the width of the
Zeeman feature. A comparison between the $p_{\rm C}$ spectra in Figs.~2 to 7 shows that the relative decrease of
the circularly polarized signal remains comparable for the unsaturated masers. However, here the weakness of the
polarization will make a detection difficult, if not impossible. The polarization monitoring
therefore has to be done with the best possible spectral resolution.
%
\onlfig{5}{
   \begin{figure}
   \includegraphics[width=6.5cm,angle=-90]{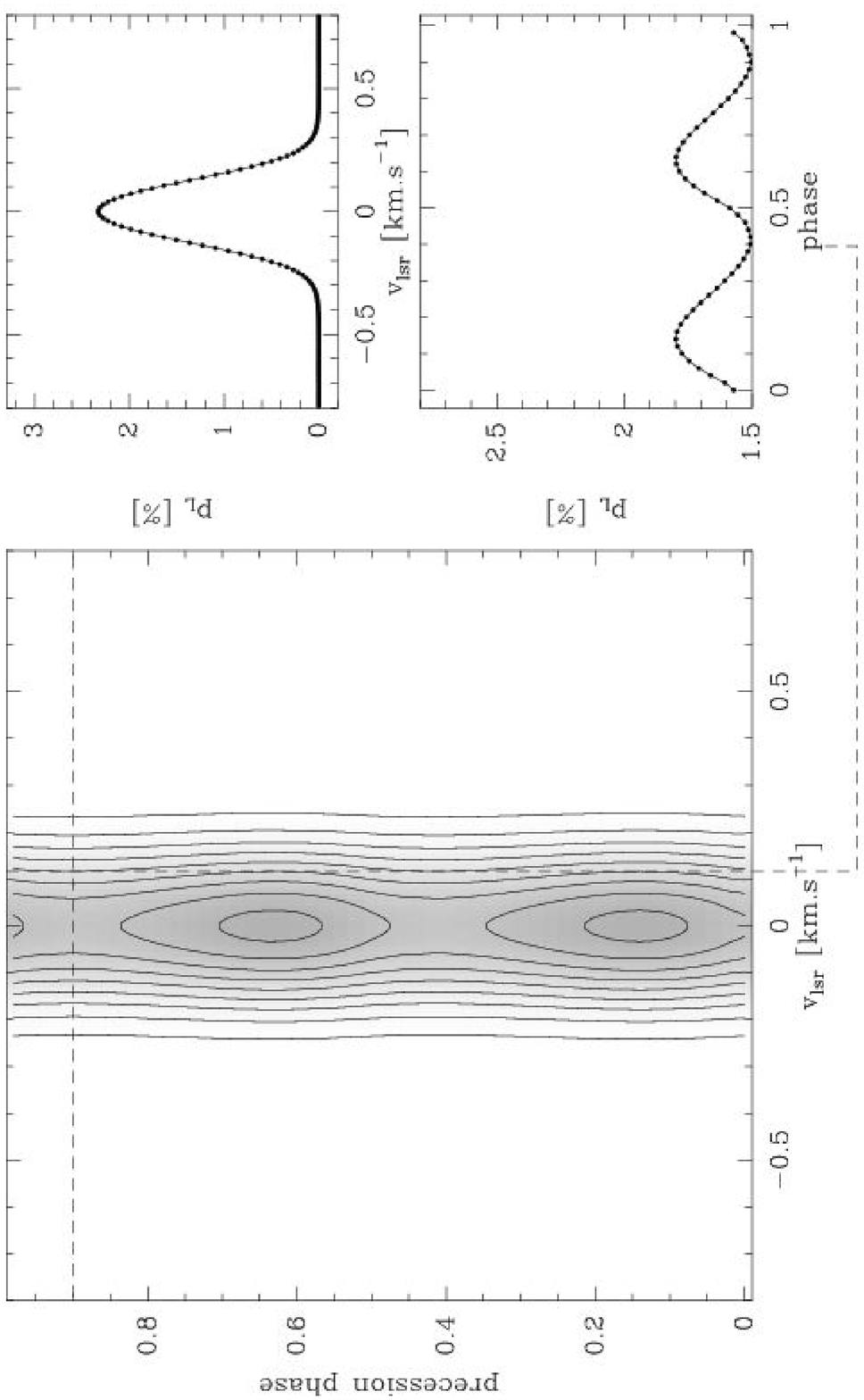}
   \includegraphics[width=6.5cm,angle=-90]{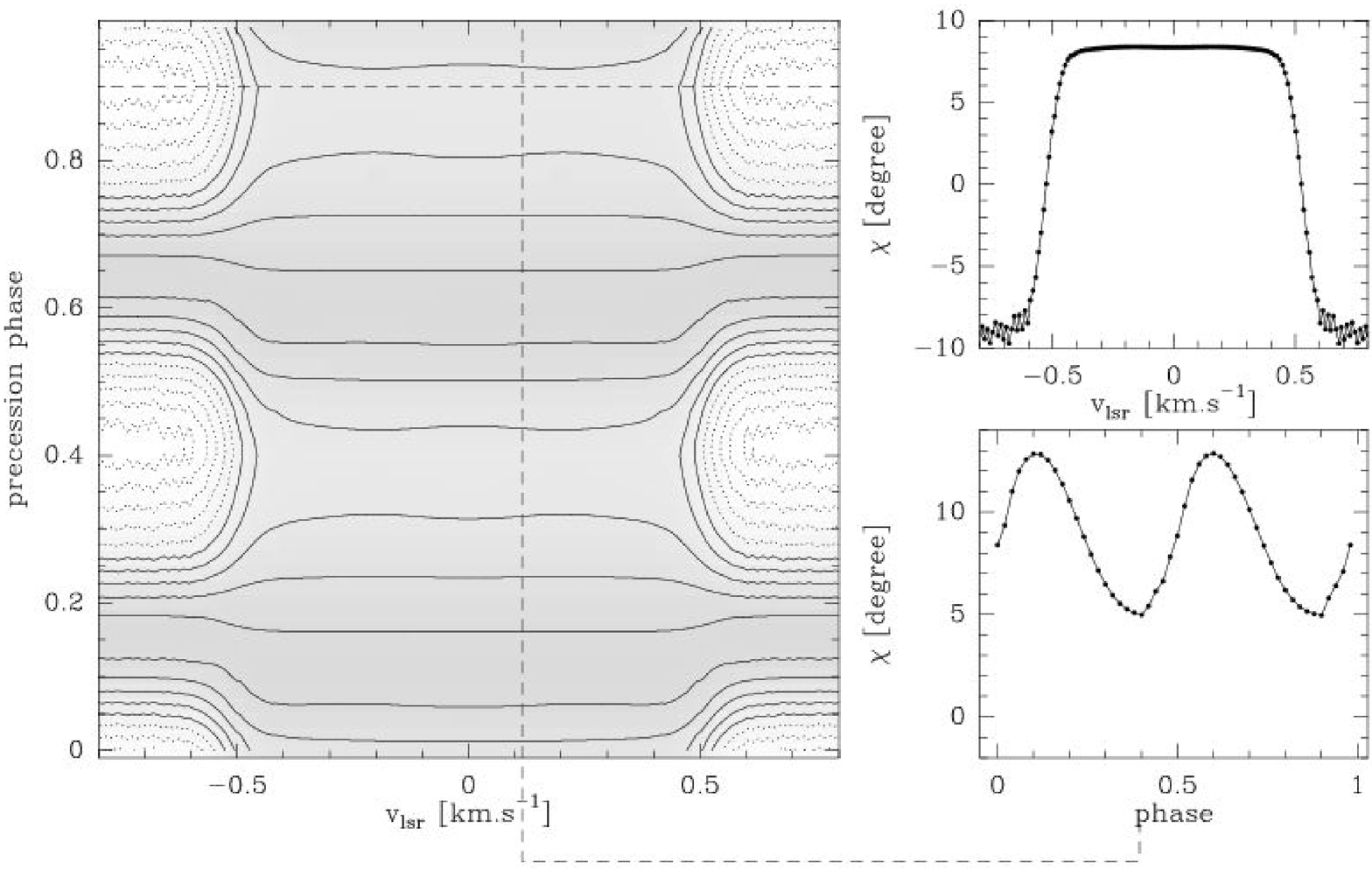}
   \includegraphics[width=6.5cm,angle=-90]{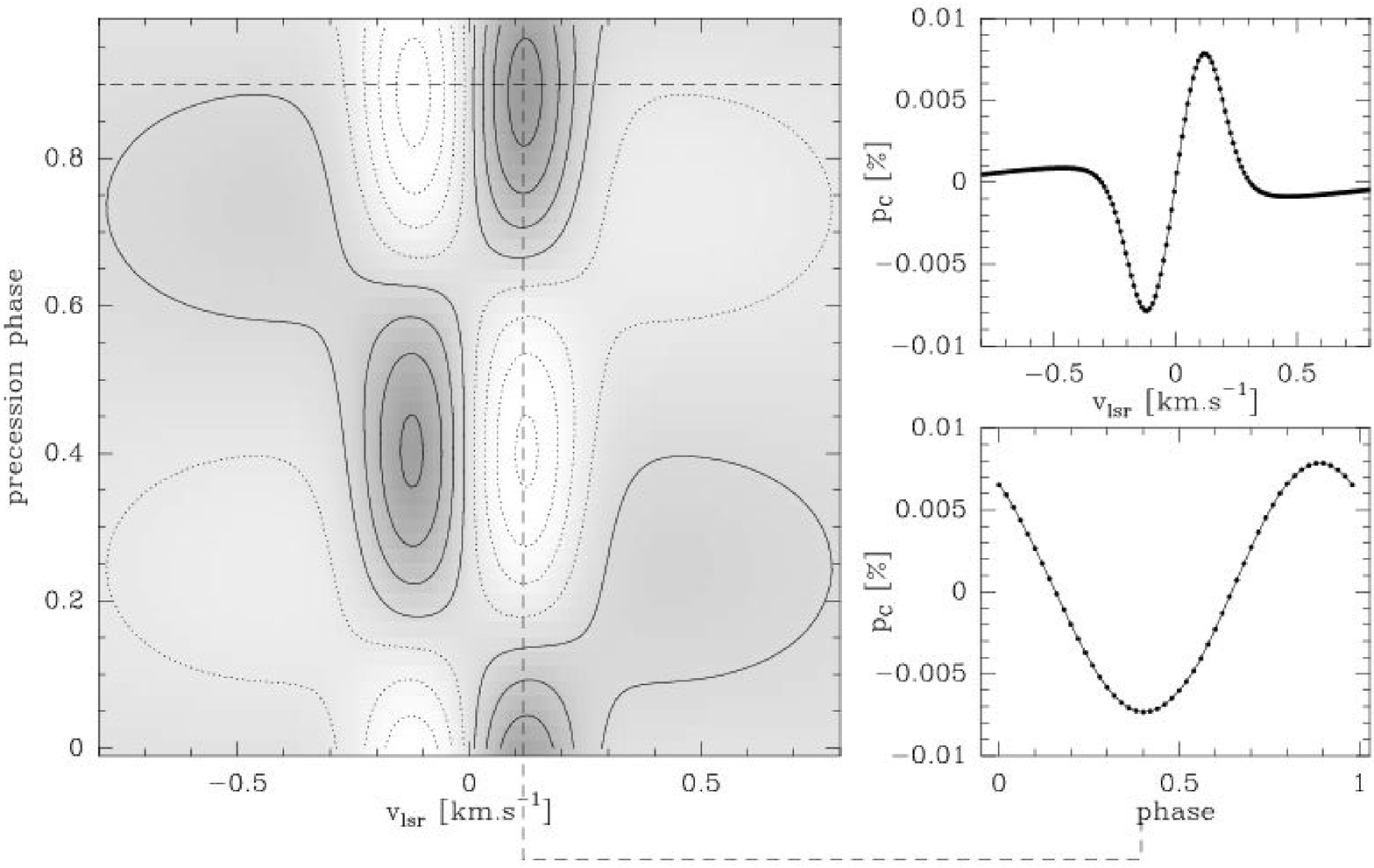}
   \caption{Same as Fig.~\ref{fig:unsaturatedPencil}, but for a bundle
            of SiO masers, with slab diameter of $3\,r_{\rm J}$ (model 
            2b). Top: $p_{\rm L}$, contour levels $0.3\,\%$ to
            $2.7\,\%$ by $0.3\,\%$. Center: $\chi$, contour levels
            from $-15^\circ$ to $+12^\circ$ by $3^\circ$. The ''noise''
            in the spectrum of $\chi$ is a numerical artifact. Bottom: 
            $p_{\rm C}$, contour levels from $-0.007\,\%$ to
            $+0.007\,\%$ by $0.002\,\%$.
            \label{fig:unsaturated}
           }
   \end{figure}
}
\section{Conclusions}
I have shown that densely sampled polarization monitoring is able to reveal precessing
Jovian magnetospheres engulfed in the atmospheres of AGB stars by a periodic modulation
of the polarization characteristics of SiO masers crossing them. The expected period is
that of the rotation of giant gas planets (i.e., $\sim 10\,$h if the rotation period
did not change with respect to typical solar system values), which allows us to distinguish
such a fingerprint from long-term variations of the polarized (Glenn et al. 2003) and
unpolarized maser flux (Humphreys et al., 2002; Diamond \& Kemball, 2003), which occur on
timescales at least an order of magnitude longer than those considered here.
As for the distinction from pseudo-periodic polarization variations due
to stellar magnetic flux loops or magnetic clouds, everything depends on the speed of the 
SiO masers relative to the latter. At a speed of 10\,km\,s$^{-1}$, a maser covers a 
distance of $5\,r_{\rm J}$ within 10\,h, i.e., the patterns modeled here are only visible if the maser 
speed across the planetary magnetosphere is $< 10$\,km\,s$^{-1}$, or if the precession
period is $< 10$\,h. Otherwise, polarization fluctuations are expected that not only
contain temporal variations, but also spatial ones, and the patterns will be either
compressed or stretched. A planetary magnetosphere exposed to an AGB wind has its
magnetopause closer to the planet than e.g., Jupiter's magnetopause, which is exposed to
the faster, but less dense solar wind. As a matter of fact, the case of a planetary
magnetosphere in an AGB wind may be closer to that of the terrestial magnetosphere,
whose magnetopause is as close as a few Earth radii, and can even be temporarily
disrupted in case of strong solar storms. Likewise, the magnetic dipole component of a
planetary magnetic field engulfed by the atmosphere of an AGB star can be expected to be
quite compact (not more than 10 planetary radii). To observe the modeled
features, an SiO maser has to cross it, and either needs to be saturated to produce a
measurable circular polarization, be unsaturated if the pump and loss rates are
sufficiently anisotropic, or has to cross a part of the magnetosphere with a strong
line-of-sight component of the magnetic field. This leads to the conclusion that the
phenomenon will be rare and, if detected, only be seen in a narrow range of velocities,
and preferentially in circular polarization-enhancing regions of strong magnetic flux
density (whereas linear polarization tends to cancel out if the polarization angle strongly
varies across the observing beam). Furthermore, the dynamics of the gas close to the
planet possibly accreting matter needs to be such that the line-of-sight velocity
coherence allows for the maser action. The abundance of SiO molecules required to form a
strong enough maser can be expected to be less of a problem. The evaporation of
Galilean moons has been proposed as an additional reservoir for the gas-phase SiO (Struck-Marcell, 
1988).

Recently, Wiesemeyer et al. (2009) have detected a pseudo-periodic variation of the fractional 
circular polarization towards two Mira stars, R~Leo, and V~Cam, which may hint at
precessing magnetospheres in their atmospheres. Several periods will have to
be monitored, to better distinguish the phenomenon from other rapid, but
transient magnetospheric events. The case of R~Leo is especially intriguing, since
VLBA maps of the 43~GHz SiO maser (Cotton et al. 2004, 2008; Soria-Ruiz et al. 2007)
show elongated SiO features pointing radially away from the star, reminiscent of jet-like
features. The velocity information from the observed polarization fluctuations, together
with the astrometry of such elongated features, suggest orbital motion in at least one of
them, and strengthens the case for a planetary wake flow as simulated by Struck et al.
(2004). Dense polarization monitoring of these features may be especially
rewarding, and a spatially fully resolved polarization monitoring will certainly be highly conclusive.
However, VLBI techniques will suffer from a confusion between polarized flux variations and variations
due to Earth rotation synthesis of visibility components (the timescale for both is unfortunately the same).
This calls for a combination of single-dish polarization monitoring with VLBI
imaging of the Stokes I emission. The cross-identification of polarization fluctuations
with VLBI features is possible if there is, in the single-dish observations, no blend of
spatially separated SiO maser spots at the same velocity or if the VLBI features show an
unambiguous association with a planetary wake flow. Detection of orbital motion in 
such features will provide a touchstone for the scenarios presented here. The aim of this paper
has been to motivate coordinated observational efforts, so far unprecedented, and to show that meaningful
results can be expected to emerge from them. If so, this will call for advanced modeling. The work
published here should therefore be considered as the initial step in an iterative refinement.

%
\onlfig{6}{
   \begin{figure}
   \includegraphics[width=6.5cm,angle=-90]{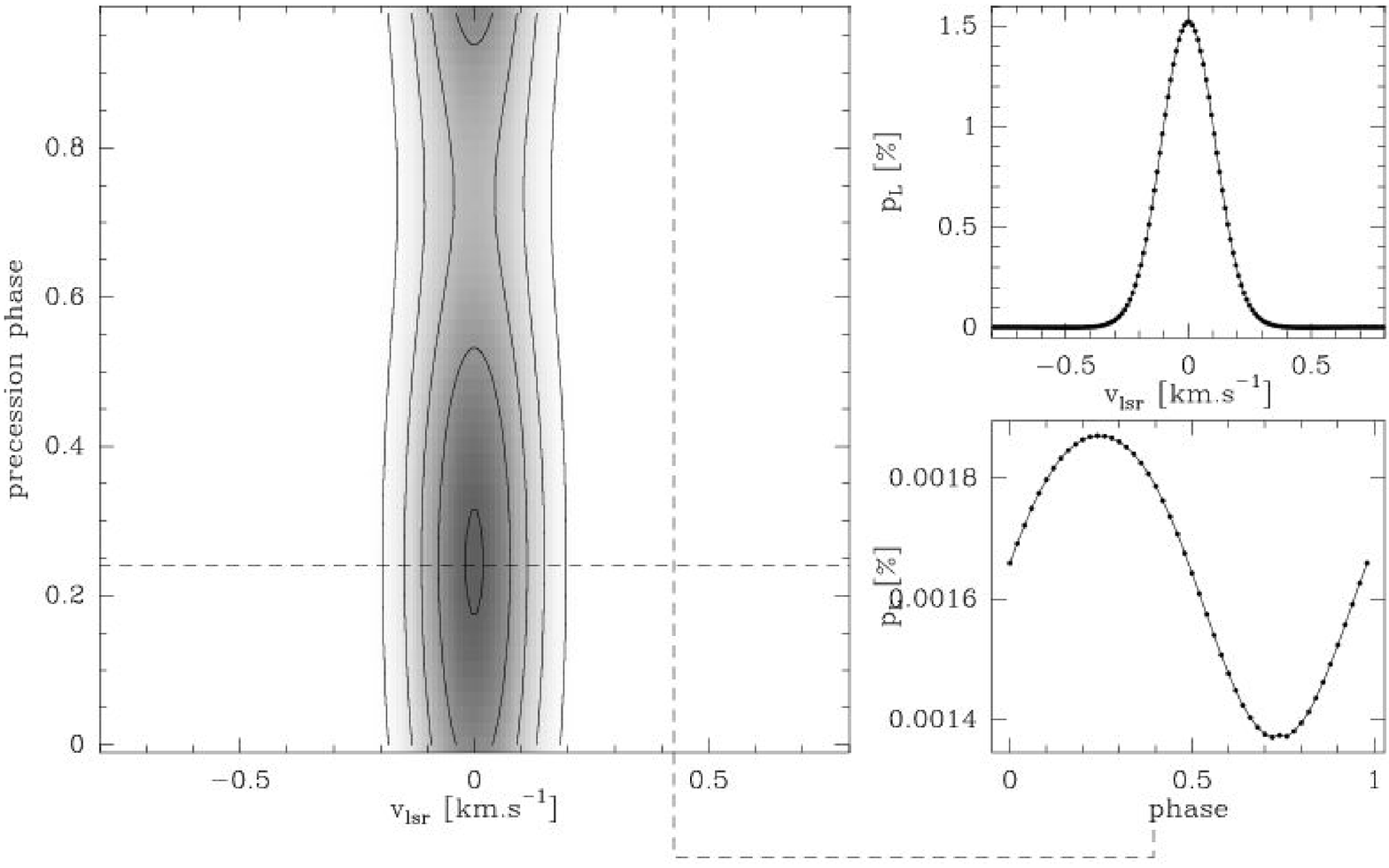}
   \includegraphics[width=6.5cm,angle=-90]{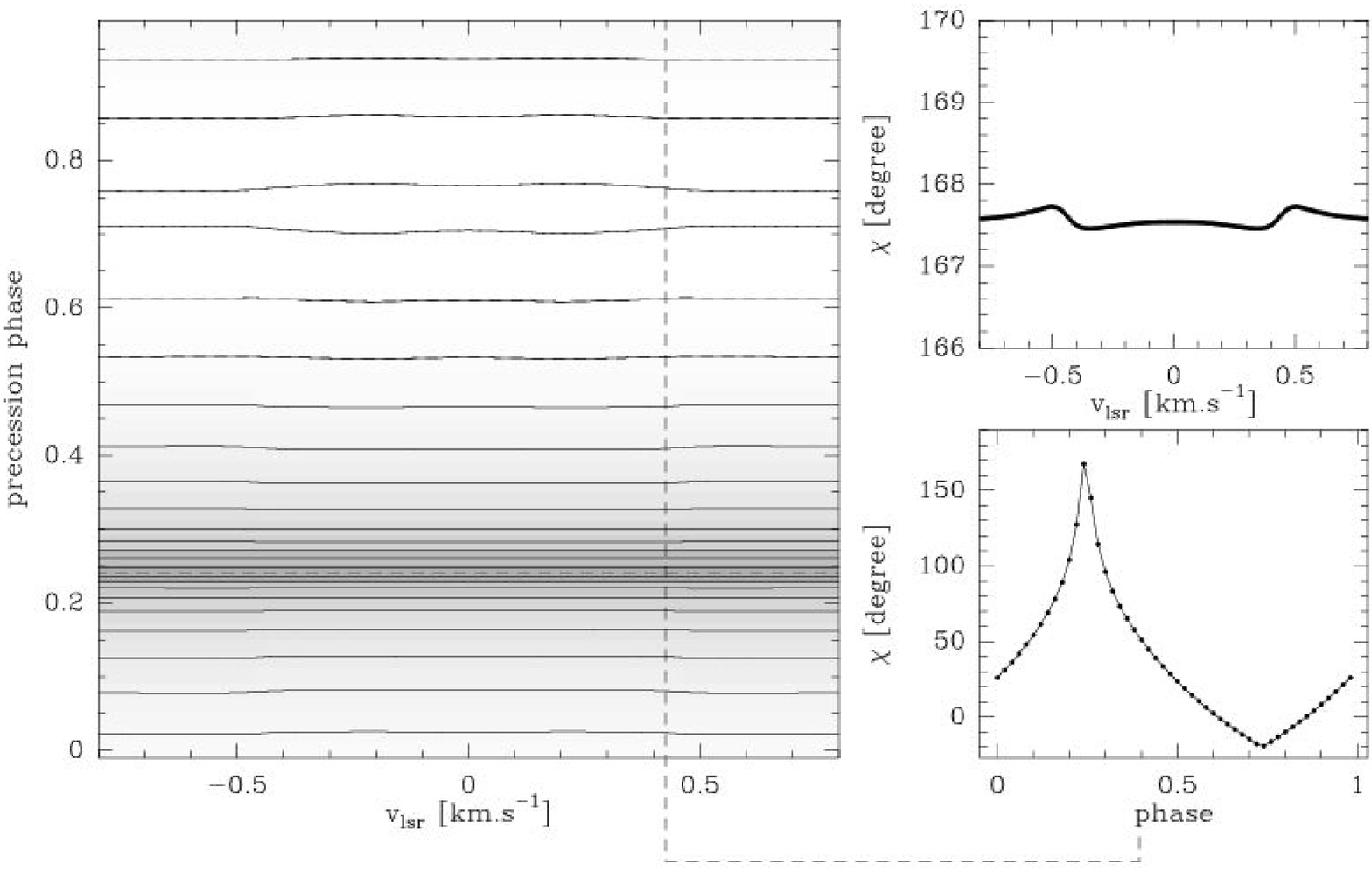}
   \includegraphics[width=6.5cm,angle=-90]{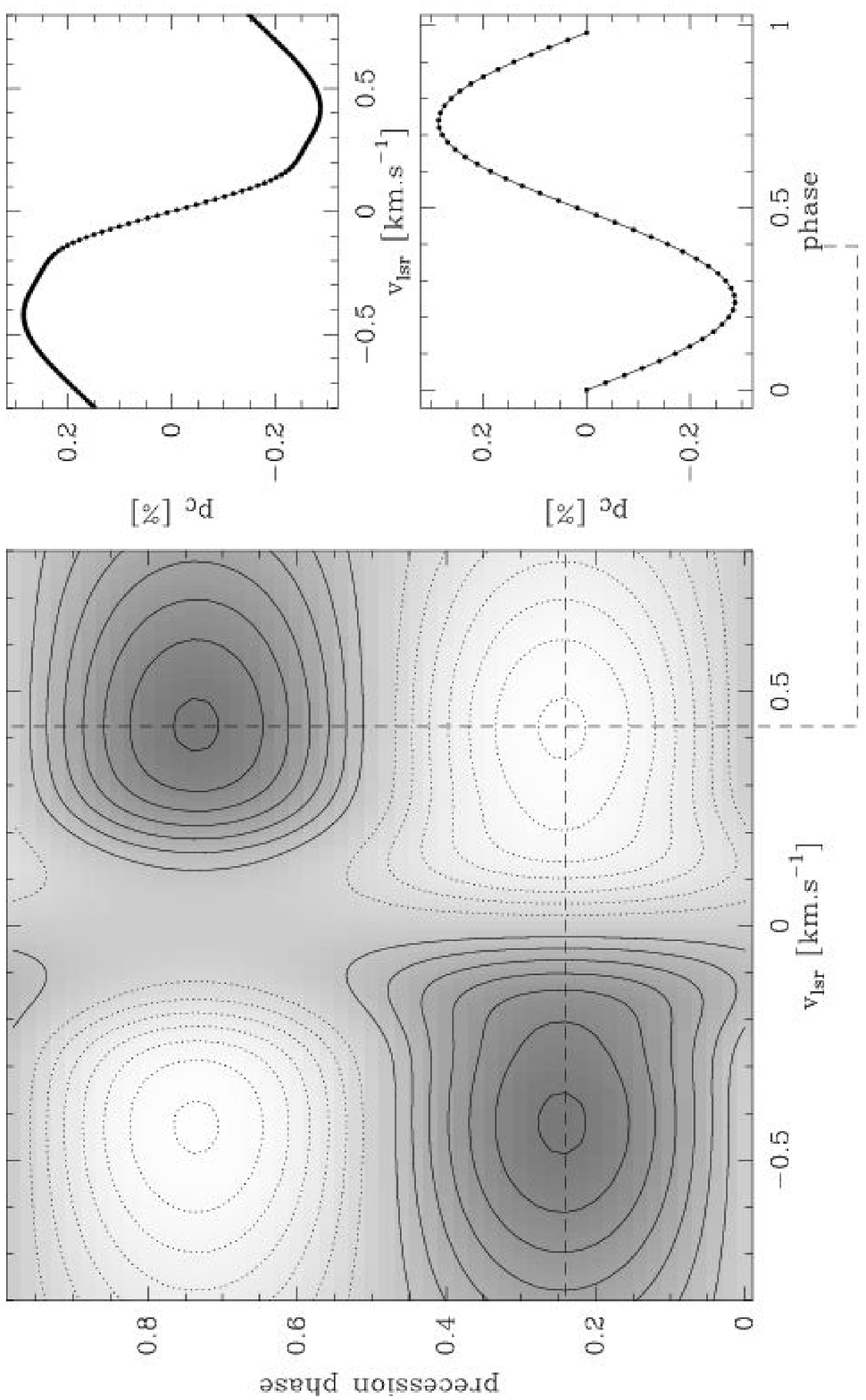}
   \caption{Same as Fig.~\ref{fig:unsaturatedPencil}, but
            $3\,r_{\rm J}$ above the planet's equatorial plane and
            towards its rotation axis (model 2c). Top: $p_{\rm L}$,
            contour levels $0.3\,\%$ to $1.5\,\%$ by $0.3\,\%$.
            Center: $\chi$, contour levels from $-16^\circ$ to
            $+160^\circ$ by $16^\circ$. Bottom: $p_{\rm C}$, contour
            levels from $-0.28\,\%$ to $+0.28\,\%$ by $0.04\,\%$.
            \label{fig:unsaturatedPencil2}
           }
   \end{figure}
}
%
\onlfig{7}{
   \begin{figure}
   \includegraphics[width=6.5cm,angle=-90]{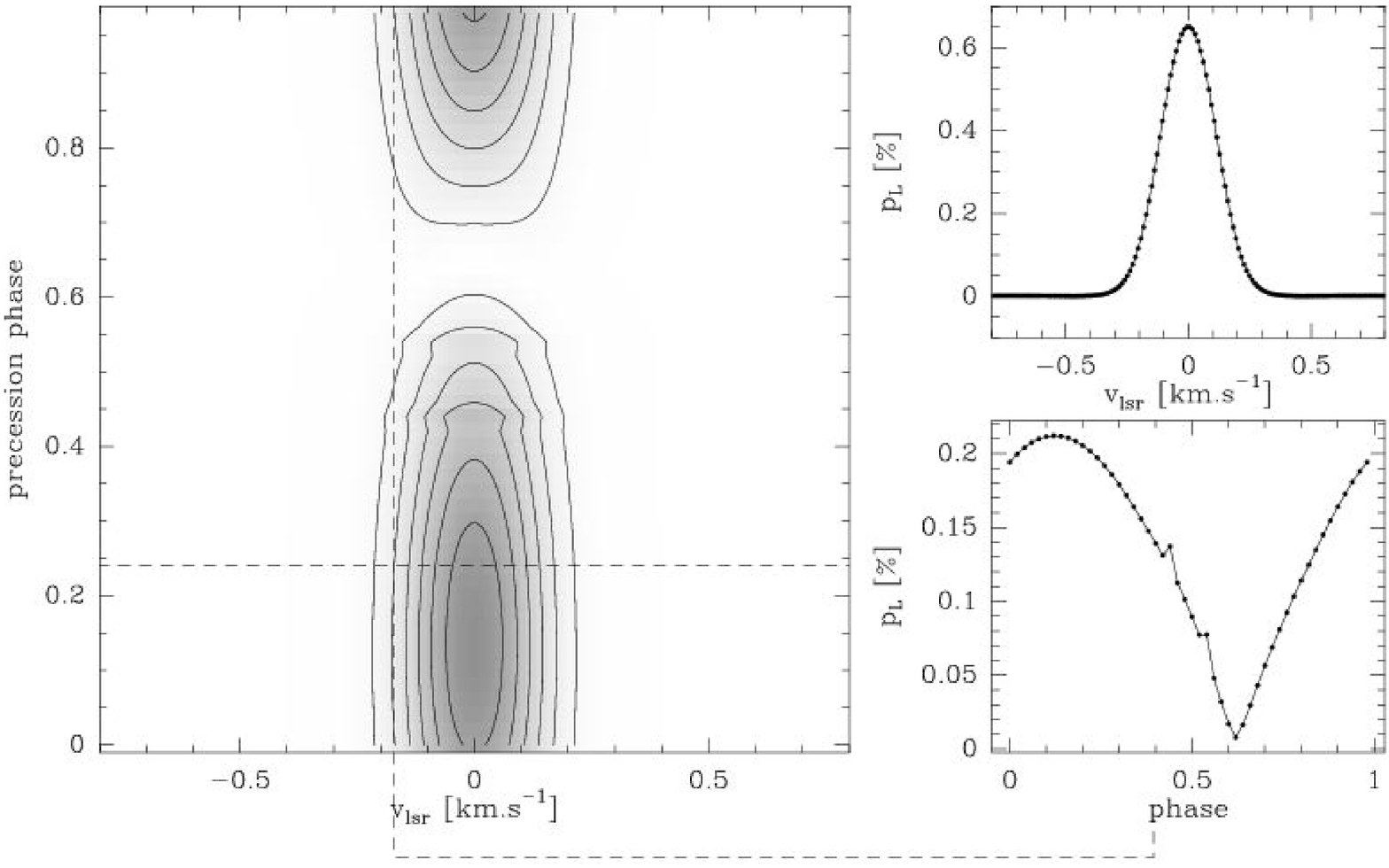}
   \includegraphics[width=6.5cm,angle=-90]{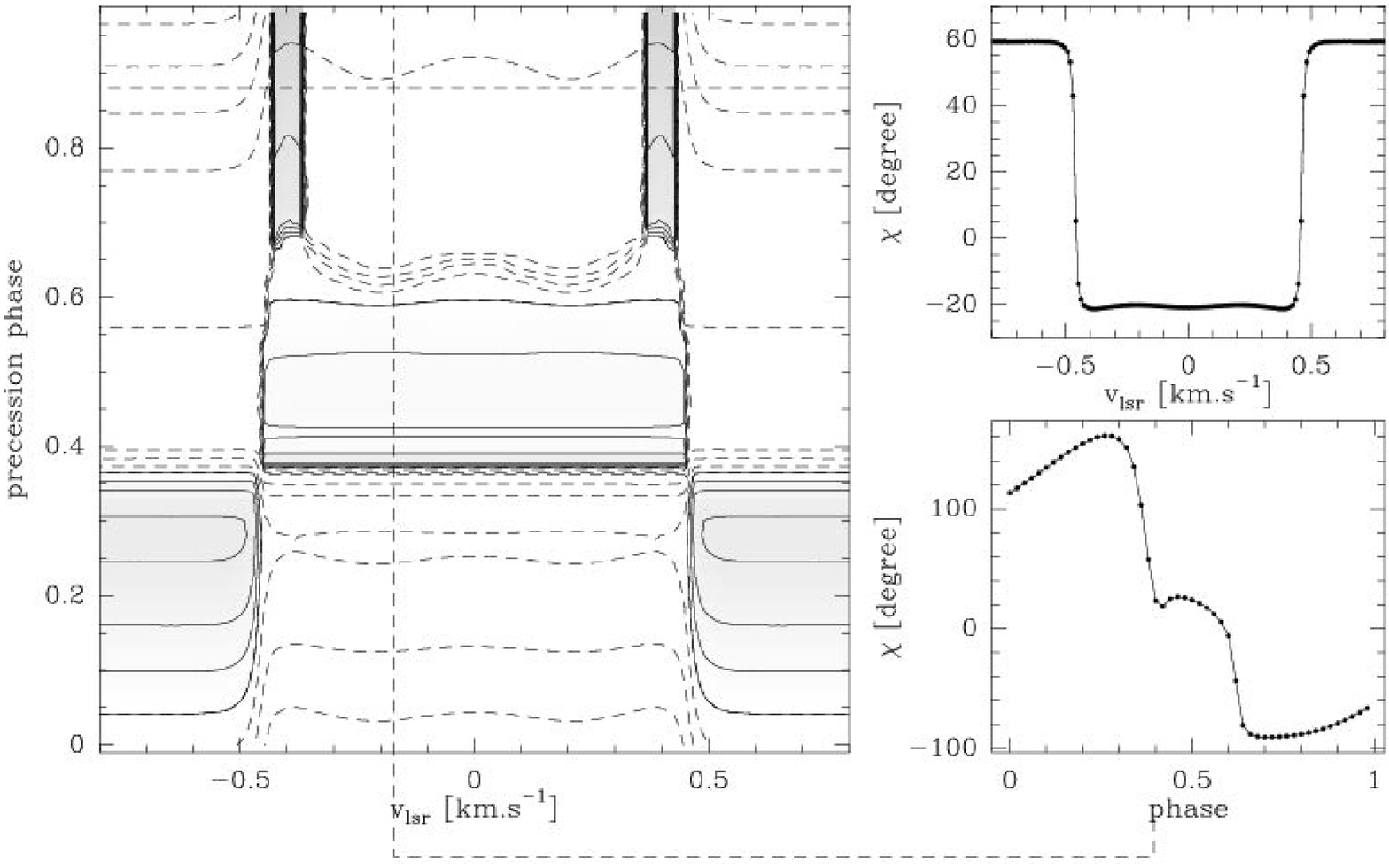}
   \includegraphics[width=6.5cm,angle=-90]{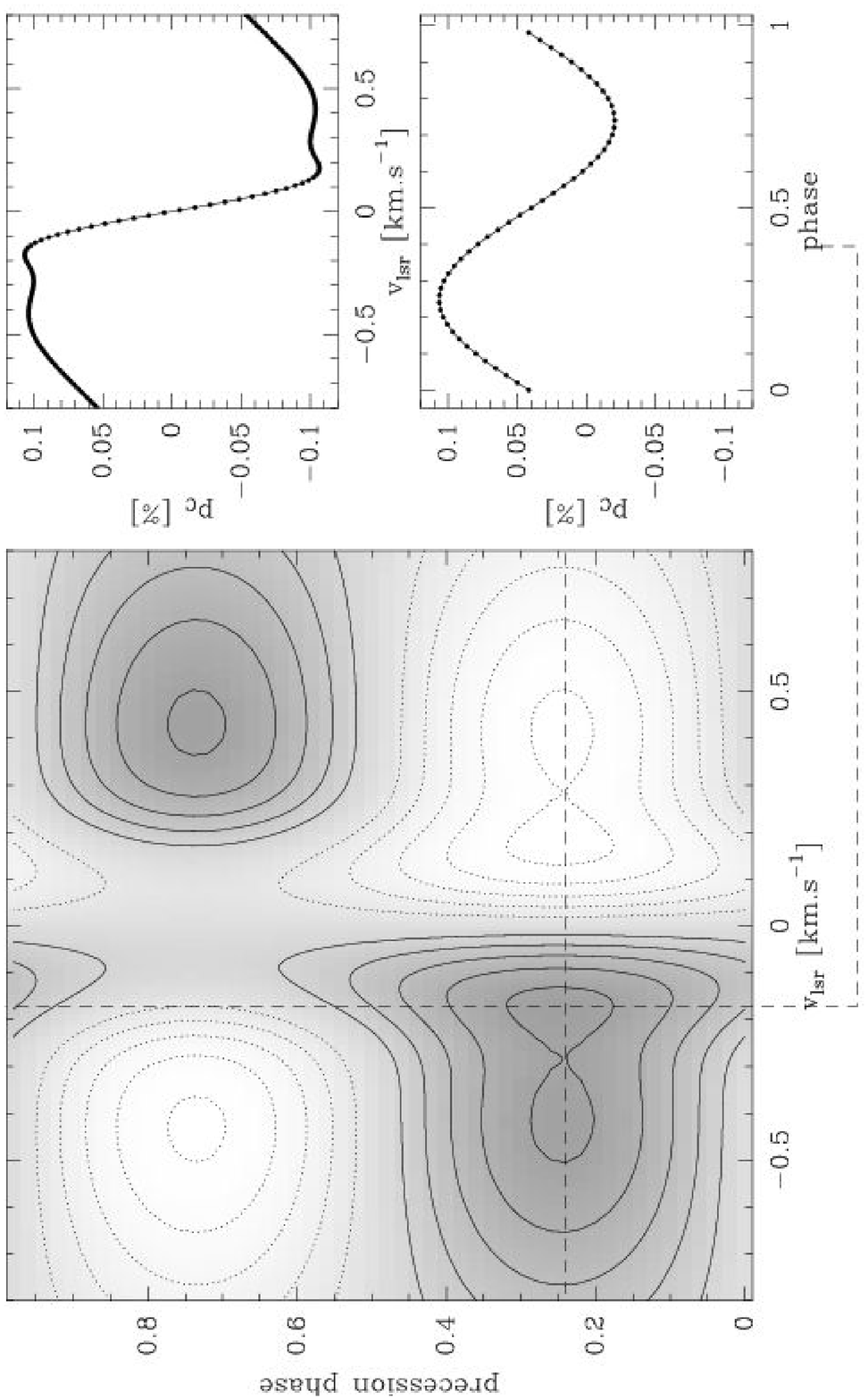}
   \caption{Same as Fig.~\ref{fig:unsaturatedPencil2}, but slightly
            offset from the rotation axis (by $0.5\,r_{\rm J}$, model
            2d). Top: $p_{\rm L}$, contour levels $0.1\,\%$ to
            $0.6\,\%$ by $0.1\,\%$ (the small jumps in the insert
            showing the time series of $p_{\rm L}$ are a numerical
            artifact). Center: $\chi$, contour levels from
            $-16^\circ$ to $+160^\circ$ by $16^\circ$. Bottom:
            $p_{\rm C}$, contour levels from $0.02\,\%$ to $+0.1\,\%$
            by $0.02\,\%$. \label{fig:unsaturatedPencil3}
           }
   \end{figure}
}
%

%

\begin{acknowledgements}
This work has been improved by the valuable comments of C.\,Thum and W.D.\,Watson,
who also helped to make the text more readable and to avoid misunderstandings.
I acknowledge the constructive comments of an anonymous referee and the assistance
of the computer group of IRAM Granada in providing the computing power for the
numerical calculations. I would like to thank J.~Adams and C.~Halliday
for their careful language editing of this and a related paper.
\end{acknowledgements}

%
\begin{appendix}
\section{The magnetosphere model}
   Before modeling a Jovian magnetosphere exposed to an AGB wind, it has to 
   be shown whether it can resist the latter, i.e., whether it is able
   to form a magnetopause far enough from the planet to allow for a
   protected magnetic dipole field, such that the ram pressure of the
   inflowing wind equals the magnetic pressure of the magnetosphere
   (Chapman-Ferraro boundary, see e.g., Kennel, 1995). This is the case if
   the magnetic flux density in the magnetopause, $B_{\rm mp}$, is
   \begin{equation}
   B_{\rm mp} = 0.05 \left( \frac{\rho_{\rm W}}{10^{-16}\,{\rm g\,cm^{-3}}}
                      \right)^{1/2}
                      \left( \frac{v_{\rm W}}{10\,{\rm km\,s^{-1}}}
                      \right) {\rm G}\,,
   \label{eq:magnetopause}
   \end{equation}
   where $\rho_{\rm W}$ and $v_{\rm W}$ are the density and velocity of the
   wind, respectively (Struck-Marcell, 1988), scaled to the typical values for
   an AGB wind. The assumed Jovian magnetic dipole field reaches this value in
   the equatorial plane at a distance
   \begin{equation}
   r_{\rm mp}  = \left( \frac{M}{B_{\rm mp}} \right)^{\frac{1}{3}}\,.
   \label{eq:distanceMP}
   \end{equation}
   For a magnetic moment of $M [{\rm G}] = 34\,r_{\rm J}^3$ (eight times
   the Jovian value), a dipole magnetosphere can extend to up to 8.8 Jovian
   radii (substellar distance from the planet). For a weaker magnetic field, 
   the magnetopause would be closer to the planet, but it would still be possible
   to have an SiO maser crossing the precessing magnetosphere in an extended
   magnetotail. 

   A comparison with the solar system may be instructive here. The Jovian
   magnetopause extends to 92\,$r_{\rm J}$, which is about double what
   would be expected from Eq.~\ref{eq:magnetopause} for solar system parameters 
   ($\rho_{\rm W} = 0.14$\,cm$^{-3}$, $v_{\rm W} = 400$\,km\,s$^{-1}$), showing 
   that the Jovian magnetosphere is more complex than a magnetic dipole field
   protected by a magnetopause (Alexeev \& Belenkaya 2005). Only a strong enough
   interplanetary magnetic field can push the magnetopause beyond
   this limit. In the atmosphere of an AGB star, even more complex phenomena
   may be expected, because of the high density of the wind and the possibility
   of a reconnection between the planetary and stellar magnetic field. For 
   comparison, the terrestial subsolar magnetopause may be as close at
   $6\,r_{\rm E}$ to the Earth's center, depending on space weather
   (Shue et al. 1997), i.e., at a distance (relative to the Earth's radius)
   rather similar to that of the Jovian model planet (relative to $r_{\rm J}$)
   than to that of Jupiter. Likewise, the location of a planetary magnetopause
   in an AGB wind depends on the variations in the latter, especially in
   response to the stellar pulsations. These timescales are, however, far longer
   than those considered here ($\sim 10$\,h). To avoid a perturbation
   of a precessing magnetosphere by the AGB wind and the accretion of matter
   onto the planet, the magnetic flux needs to be decoupled from the matter 
   via ambipolar diffusion. As for the latter, I estimate its speed
   $v_{\rm ad}$ following Hartquist \& Dyson (1997), but using values closer
   to the magnetosphere modeled here, by
  \begin{eqnarray}
   v_{\rm ad} & \simeq 1000\, {\rm km\,s^{-1}} & \left(\frac{B}{1\,\rm G}\right)^2
   \left(\frac{L}{0.1 \, r_{\rm J}}\right)^{-1}
   \left(\frac{X_{\rm i}}{10^{-6}}\right)^{-1} \nonumber \\
              &        & \cdot \left( \frac{n_{\rm H}}{10^{10}\,
    {\rm cm^{-3}}}\right)^{-2}
   \label{eq:ad}
   \end{eqnarray}
   where $L \simeq 0.1\,r_{\rm J}$ is the typical length scale on which the
   modeled magnetic field varies, $B \simeq 1$\,G is the typical magnetic flux
   density in the model magnetosphere, $X_{\rm i} \simeq 10^{-6}$ the ionization
   fraction, and $n_{\rm H} \simeq 10^{10}\,{\rm g\,cm^{-3}}$ the expected 
   hydrogen density in the maser slab. The major uncertainty in Eq.~\ref{eq:ad}
   comes from the gas density (especially if shocks are involved) and the
   ionization fraction. In any realistic case, the suggested ambipolar
   diffusion speed of $\sim 1000$\,km\,s$^{-1}$ is much faster than the
   precession speed of the magnetosphere
   ($2\pi r_{\rm J} /T \sim 12$\,km\,s$^{-1}$ with a period
   of $T = 10$\,h), and the magnetic flux and matter can decouple efficiently,
   such that the magnetic field lines diffuse through the maser slab.
   This conclusion holds for the localized magnetic field considered here. As
   for the global magnetic field pervading the AGB atmosphere, Soker (2006)
   rules out any dynamical importance, because of the weak coupling between the
   magnetic flux and the matter, based on a similar argument. I conclude that
   a planetary magnetic dipole field can persist in the extended atmosphere of an AGB star.
   Ohmic dissipation of the magnetic flux can be neglected here, otherwise the
   substantial circular polarizations of SiO masers could not have been
   observed. 
\section{Numerical calculations}

   Because the growth rates of the Stokes parameters in the propagation of
   maser photons differ by orders of magnitude at different velocities, Eq.~7
   is a stiff set of differential equations, and numerical solutions using
   Runge-Kutta or multiple-order methods are inefficient, even with adaptive
   step sizes. Because the Stokes parameters at different frequencies are not 
   independent of each other due to the Zeeman splitting, the
   calculations have to be done along the full zero-to-zero width
   of the spectral line, with a sampling dense enough to resolve
   the Zeeman feature. For a given time step $t$, the calculations
   proceed along the line-of-sight in steps that are short enough to solve
   the equation of radiative transfer using the matrix exponential function
   $\exp{({\mathbf K}\tau)}$ with piecewise constant coefficients, i.e.,
   for a step from grid point $j$ to $j+1$
   \begin{equation}
        {\bf I}_{\rm j+1} = \exp{({\bf K}\tau)} {\bf I}_{\rm j}
   \label{eq:matrixExponential}
   \end{equation}
   where the elements of the vector ${\bf I}_{\rm j}$ are the Stokes parameters
   entering the grid cell. Spontaneous emission is neglected, as usual in maser theory.
   Landi degl'Innocenti \& Landi degl'Innocenti (1981) have provided
   an elegant analytical solution for the matrix exponential function 
   in closed form. At each sampling point along the line-of-sight, we
   calculate the magnetic field from Eqs.~$19-23$, the corresponding
   angles $\gamma$ and $\eta$ from Eq.~\ref{eq:gammaEta}, the
   resulting Zeeman splitting $g\Omega$, and the corresponding
   matrix elements $A$ to $E$. For a given line-of-sight element and frequency,
   the stimulated emission rates in Eqs.~1 to 5 are
   determined from the normalized Stokes intensities entering it
   by a Spline interpolation of the spectral line profiles in
   $(I,Q,U,V)^{\rm T}$. I assume a linear, unidirectional maser,
   with unpolarized continuum seed radiation. For each rotation
   phase considered, results are shown for both a maser spot much
   smaller than the length scale on which the magnetic field 
   typically varies, and for a maser spot consisting of a number
   of independent, parallel rays (hereafter called {\it bundle}). As
   already mentioned, for distances farther than $8\,r_{\rm J}$ away from 
   the planet the model would not be valid anymore, and would have to be
   modified by the shocked AGB wind, the magnetopause, and the
   magnetotail. Without a detailed magneto-hydrodynamical model at hand, we
   assume that all masers of the bundle are centered on an impact
   parameter of $3\,r_{\rm J}$ (such that the maser is confined within
   the magnetic dipole field), have the same line-of-sight velocity, and are
   characterized by a total gain length of $10\, r_{\rm J}$. The tidal
   interaction of the masing gas with the possibly matter accreting planet
   leads to a loss of velocity coherence and thus of maser saturation. I
   therefore model the effects of an incoherent velocity field by comparing
   unsaturated maser action with a saturated one. For the case of a homogeneous 
   magnetic field the code has been successfully tested by comparing its
   results whith those of Watson \& Wyld (2001).
\end{appendix}

\end{document}